\documentclass[12pt,preprint]{aastex}
\usepackage{apjfonts}
\usepackage{amssymb, amsmath, amsbsy, booktabs} %, epsfig, epsf, rotating}
\usepackage{threeparttable}
\usepackage[dvips]{color}
\slugcomment{\today}
\shorttitle{SDSS~J0838$+$5406: IMBH host galaxy with twenty kpc-scale \ha\ emission}
\shortauthors{W.-J.~Liu et al.}

\def\flux{erg~s$^{-1}$~cm$^{-2}$}

\newcommand{\msun}{\ensuremath{M_{\odot}}}

\newcommand{\ergs}{\ensuremath{\mathrm{erg~s^{-1}}}}
\newcommand{\kms}{\ensuremath{\mathrm{km~s^{-1}}}}
\newcommand{\mbh}{\ensuremath{M_\mathrm{BH}}}

\newcommand{\chisq}{\ensuremath{\chi^2}}

\newcommand{\ha}{{\rm H\ensuremath{\alpha}}}
\newcommand{\hb}{{\rm H\ensuremath{\beta}}}

\newcommand{\hd}{H\ensuremath{\delta}}

\newcommand{\hi}{H\,{\footnotesize I}}
\newcommand{\hii}{H\,{\footnotesize II}}
\newcommand{\nii}{[N\,{\footnotesize II}]}
\newcommand{\sii}{[S\,{\footnotesize II}]}
\newcommand{\oiii}{[O\,{\footnotesize III}]}

\newcommand{\oii}{[O\,{\footnotesize II}] $\lambda$3727}
\newcommand{\oi}{[O\,{\footnotesize I}]}

\newcommand{\hst}{\textit{HST}}
\newcommand{\etal}{et~al.}
\newcommand{\thisobj}{J0838$+$5406}

\newcommand{\rev}[1]{{#1}}  %%2017Feb3

\begin{document}

\title{A Ringed Dwarf LINER 1 Galaxy Hosting an Intermediate-mass Black Hole with Large-scale Rotation-like \ha\ Emission}
%% SDSS\,J083803.68$+$540642.0: ---

\author{Wen-Juan~Liu\altaffilmark{1,2},  Lei Qian\altaffilmark{3,2},
Xiao-Bo~Dong\altaffilmark{1,2}, Ning~Jiang\altaffilmark{4}, Paulina~Lira\altaffilmark{5},\\
Zheng~Cai\altaffilmark{6}, Feige~Wang\altaffilmark{7}, Jinyi~Yang\altaffilmark{7},
Ting~Xiao\altaffilmark{8}, and Minjin~Kim\altaffilmark{9,10} }

\altaffiltext{1}{Yunnan Observatories, Chinese Academy of Sciences, Kunming, Yunnan 650011, China;
Key Laboratory for the Structure and Evolution of Celestial Objects, Chinese Academy of Sciences,
Kunming, Yunnan, 650011, China;
 \mbox{wjliu@ynao.ac.cn} , \mbox{xbdong@ynao.ac.cn} }
%% \mbox{wjliu@ynao.ac.cn}  }
\altaffiltext{2}{Center for Astronomical Mega-Science, Chinese Academy of Sciences,
20A Datun Road, Chaoyang District, Beijing, 100012, P. R. China}
\altaffiltext{3}{National Astronomical Observatories, Chinese Academy of Sciences, Beijing, 100012, China}
\altaffiltext{4}{Department of Astronomy, University of Science and Technology of China, Hefei, Anhui 230026, China}
\altaffiltext{5}{Departamento de Astronom\'ia, Universidad de Chile, Casilla 36-D, Santiago, Chile}
\altaffiltext{6}{UCO/Lick Observatory, University of California, 1156 High Street, Santa Cruz, CA 95064, USA}
\altaffiltext{7}{Department of Astronomy, School of Physics, Peking University, Beijing 100871, China}
\altaffiltext{8}{Key Laboratory for Research in Galaxies and Cosmology, Shanghai Astronomical Observatory, Chinese Academy of Sciences,
80 Nandan Rd, Shanghai 200030, China}
\altaffiltext{9}{Korea Astronomy and Space Science Institute, Daejeon 305-348, Korea}
\altaffiltext{10}{University of Science and Technology, Daejeon 305-350, Korea}

%%%% begins the real content:  %%%%%
%%%%
\begin{abstract}
  We report the discovery of a 20-kpc-sized \ha\ emission in SDSS\,J083803.68$+$540642.0, a
  ringed dwarf galaxy ($M_{V} = -17.89$ mag) hosting an accreting intermediate-mass black hole at $z = 0.02957$.
  Analysis of the \hst\ images indicates that %%\thisobj\
  it is an early-type galaxy with a featureless low-surface brightness disk
  (\mbox{$\mu_0 = 20.39$\,mag~arcsec$^{-2}$} in the $V$ band)
  %% disk --2016Nov19 %%
  and a prominent, relatively red bulge ($V - I$ = 2.03, $R_{\rm e} = 0.28$\,kpc or 0\farcs48)
  that accounts for $\approx$81\% of the total light in the $I$ band.
  A circumgalactic ring of a diameter 16~kpc is also detected, with a disperse shape on its south side.
  The optical emission lines reveal the nucleus to be a broad-line LINER.
 Our MMT longslit observation indicates that
the kinematics of the extended \ha\ emission is consistent with a rotational gaseous disk,
with a mean blueshifted velocity of
 162~\kms\ and mean redshifted velocity of 86~\kms.
According to our photoionization calculations, the large-scale \ha\ emission is unlikely to be powered by the central nucleus
%% 2016Nov19---
or by hot evolved (post-AGB) stars interspersed in the old stellar populations,
%% --- above, 2017Jan17.
but by {\it in situ} star formation;
this is vindicated by the line-ratio diagnostic of the extended emission.
  We propose that both the ring and large-scale \ha-emitting gas are created by the tidal accretion in a collision---and then
  merger---with a gas-rich galaxy of a comparable mass.
\end{abstract}


\setcounter{footnote}{0}
\setcounter{section}{0}

\section{Introduction}

%% first of all, introducing the common viewpoint about IMBH AGNs , in local dwarf galaxies
%% paragraph 1:
 It has been generally believed (see \citealt{2013ARA&A..51..511K}, \citealt{2016ASSL..418..431K} for recent reviews) that,
unlike massive galaxies at high redshifts (e.g., $z\gtrsim2$) whose evolution is driven by major merger,
 low-redshift galaxies largely evolve through secular evolution (i.e., in a slow and gentle manner)
 driven by internal processes within galactic disks
 and/or by environmental effects
 such as harassment or nurturing (see \citealt{2004ARA&A..42..603K} for details; see also
 \citealt{1996AJ....112..839C}, \citealt{2014MNRAS.444.1125C} for the cosmic evolution).
 %% 2016Nov19---: %%
It has also been generally believed accordingly that
most activity (namely active galactic nuclei---AGNs)
 of supermassive black holes (BHs) at the  %%center
 centers of galaxies at low redshifts
 (e.g., $z\lesssim 1$, \citealt{2011ApJ...726...57C};
 and even up to $z\sim 2$, \citealt{2012ApJ...744..148K})
 appear to be fueled by the random accretion of gas via internal, secular processes
 working close to the BH (say, within a few hundred parsec, \citealt{2013ApJ...776...50C};
 also cf. \citealt{2006ApJS..166....1H}).
In these cases there is no connection between AGN activity and major mergers of their host galaxies.
This non-connection appears particularly true for low-$z$ AGNs hosting intermediate-mass black holes (IMBHs)%
\footnote{\,Following \citet{2007ApJ...670...92G} and \citet{2012ApJ...755..167D}, hereinafter we refer to BHs with
$\mbh < 2 \times 10^6$~\msun\ at the centers of galaxies as ``low-mass'' or
``intermediate-mass'' BHs;
accordingly, for the ease of narration wherever it is not ambiguous,
hereinafter we call AGNs hosting low-mass BHs
as low-mass AGNs or IMBH AGNs.
Normally we prefer ``intermediate-mass BHs'' to ``low-mass BHs''
because of the possible confusion of the latter with the stellar-mass BHs in low-mass X-ray binaries (LMXBs).}:  %
according to the analysis of their images by Hubble Space Telescope (\hst),
the majority of low-mass AGNs live in late-type disk galaxies without a classical bulge
\citep{2008ApJ...688..159G,2011ApJ...742...68J}.
Indeed, the accretion rate for an IMBH is so tiny
($<$0.05 \msun~yr$^{-1}$ even if at the maximum Eddington accretion)
that a steady supply of fuel, in the form of stellar mass loss from evolved stars or Bondi accretion of hot gas
in the innermost regions,
is available readily much more than required \citep{2008ARA&A..46..475H}.
``In fact, the paradox for local BHs is not whether there is enough fuel to light them up.
Rather, the puzzle is how to keep them so dim despite the ready abundance of {\it in situ} gas.''
(\citealt{2013ARA&A..51..511K}; see also \citealt{2008ARA&A..46..475H}).
%%http://adsabs.harvard.edu/abs/2008ARA%26A..46..475H (Ho 2008)
That is, the real problem seems to be this \citep[cf.][]{2009ApJ...699..626H}: there must be some mechanism (yet to know)
to hinder the innermost fueling process.%
%%http://adsabs.harvard.edu/abs/2009ApJ...699..626H
\footnote{\,\rev{According to W.-M. Gu (2016, private communication),
 AGN outflows, which can be launched even at very low accretion rates (e.g.,
 Wang \etal\ 2013, Gu 2015), provide such a mechanism. }
}%%endoffootnote %%2017Jan24

%% then, Surprise!  -- paragraph 2:
%%
On the other hand, surprisingly
\citet{2011ApJ...742...68J} reported that 9\% of their IMBH host galaxies are detected
 in interacting systems with close companions (of comparable mass)
 by simple visual identification of their \hst\ images.
This fraction is already much higher than that for
 local luminous galaxies (see their \S5.2 for the detail).%
 \footnote{\,As \citeauthor{2011ApJ...742...68J} cautioned (``we identify the companions only visually based on our \hst\ images.
 We do not use well-defined criteria, ...as this is not our main focus.''),
 they only identified ongoing mergers (and soon-to-be mergers) of two or more
 galaxies of comparable masses (see their Figure~4),
 and did not take into account any morphological features indicative of {\em past} major mergers or
 other violent interactions.}
 Likewise, we inspected the SDSS images of the IMBH AGNs at $z<0.15$ in the \citet{2012ApJ...755..167D} sample,
and found that 41 (16.2\%) of the 253 objects are undergoing (dwarf--dwarf) major mergers,
minor mergers with a larger galaxy (i.e., violently impacted by the primary),
or collisions (violent encounters) with comparable or more massive galaxies,
or have features (tidal tails, shells, collisional rings, etc.) indicating recent violent interactions
of the above-listed three cases (W.-J.~Liu~\etal\ 2017, in preparation).
Such a high incidence of {\em violent} galaxy interactions
(versus internal and environmental secular processes) in IMBH AGNs
is higher than previously deemed in the literature \citep[cf.][]{2013ARA&A..51..511K}.

%% theoretically possible , contrary to the Kormendy & Ho 2013:
%%% so, it's interesting to explore this subject. -----
From a theoretical perspective, in the hierarchical framework of galaxy formation
major mergers between dwarf galaxies are not uncommon in the low-$z$ Universe,
let alone at earlier times.
For instance,  by means of cosmological simulations
\citet{2014ApJ...794..115D} found that 10\% of satellite dwarf galaxies with stellar mass $M_\star >  10^6$\msun\
within the virial radius of a M31-like host dark matter (DM) halo
experience a major merger since $z = 1$,
and furthermore that for dwarf galaxies that are outside of the host virial radius,
the major-merger frequency  doubles.
%%%2016Nov26: after Caiz's comment ---
%Importantly, recent observational evidence for it has been mounting
%\citep[e.g.,][]{2005AJ....129.2617G,2013ApJ...779L..15N,2014ApJ...795L..35C,2014Natur.507..335A,2015AJ....149..114P}.
Recently such a considerable occurrence of nearby dwarf--dwarf major mergers
has been supported by observations
\citep[e.g.,][]{2005AJ....129.2617G,2013ApJ...779L..15N,2014ApJ...795L..35C,2014Natur.507..335A,2015AJ....149..114P}.
Provided that the effect of major mergers to AGN activity is similar
in low-$z$ dwarf galaxies,
as in building-block galaxies of a similar stellar mass in the early Universe
or as in massive galaxies at $z=2-3$ supposedly triggering quasars \citep[see][]{2013ARA&A..51..511K},
we can expect that major mergers and other violent interactions of
IMBH host galaxies
(i.e., interactions transforming the configurations of the stars and disturbing the gas of the galaxies
forcefully and rapidly)
would cause the aforementioned hindrance in the inner regions (whatever it may be) disappear
and thus enable or augment the AGN fueling.

%% revised 2016.9.17 night, stop here.

%% So, we need to do it!
%%%%
Thus it is worth launching a systematic  observational study to investigate the outskirts and neighborhood
of the IMBH host galaxies, and the extent, morphology and kinematics of
cold (\hi\ and CO; L.Qian~\etal\ 2017, in preparation) and warm gas,
and  thoroughly
explore the possible correlations
of IMBH accretion activity with (violent) galaxy interaction and/or galactic environment.
As a start of our study, in this paper we present detailed observations of an individual source,
SDSS~J083803.68$+$540642.0 (hereafter \thisobj), which has been observed by the MMT telescope
with longslit spectroscopy (\S2)
and has archival \hst\ imaging data (\S3).
\thisobj\ caught our attention because it is among
the ten ``LINER~1'' (type-1 AGN of Low-ionization nuclear emission-line region; \citealt{1997ApJ...487..568H})
galaxies in the \citet{2012ApJ...755..167D} sample of 309 low-$z$ IMBH AGNs (see also \S4),
and because its optical spectrum is very red ($u-r = 2.4$; cf. \citealt{2007ApJS..173..342M})
%% Martin DC, Wyder TK, Schiminovich D, et al. 2007. Ap. J. Suppl. 173:342
as dominated by starlight of old age (see also \S5).
It belongs to the one third galaxies in the \citet{2012ApJ...755..167D} sample
with the 4000\AA\ break $D_{4000} > 1.5$,
i.e., with a mean stellar age greater than 1~Gyr (\citealt{2003MNRAS.341...33K}).
%%($\sim$2.5~Gyr according to the 4000\AA\ break $D_{4000}=1.7$).
Later on we took the MMT spectroscopic observation, and
surprisingly the 2-dimensional longslit spectrum reveals rotation-like \ha\ emission extending along the
spatial axis visually of 32\arcsec\ across (namely a projected physical size of 19.4~kpc),
which is likely powered by {\it in situ} star formation (\S6).
Then we searched archival data and found its aforementioned \hst\ images.
The images reveal a spectacular ring around the galaxy with a diameter of 15.6~kpc,
which we argue should be %%proves to be  %%2017Jan24
created by tidal accretion in a collision (and then merger)
with a galaxy of a comparable mass (see \S7).
%%% 2017Jan24:
%% Thus \thisobj, presented here as a case study,
%% provides direct evidence for the importance of violent galaxy interactions
%% plausibly even for IMBH AGNs and their host galaxies.
A brief summary and future work are given in \S8.

Throughout the paper we assume a cosmology with $H_{0} =70$ km~s$^{-1}$~Mpc$^{-1}$,
$\Omega_{\rm m} = 0.3$, and $\Omega_{\Lambda} = 0.7$.

%%// end of the revised Introduction. -- 2016Sep18 --

%\section{Data analysis}

\section{Optical spectroscopic observation and analysis}

%% to overview the SDSS spectrum first:
 \thisobj\ was in the spectral data set of SDSS DR4.
 It was compiled into the IMBH AGN samples of \citet{2007ApJ...670...92G} and \citet{2012ApJ...755..167D},
because a faint broad \ha\ line was identified inferring a viral BH mass  below 2$\times10^{6}$\msun\
after careful handlings of starlight subtraction and emission-line deblending.%
\footnote{\,It was in the `$c$' (candidate) sample of \citet{2007ApJ...670...92G}, i.e.,
not satisfying their quantitative detection threshold of broad \ha\ but being picked up manually. }
%%  and was classified as a galaxy by the SDSS pipeline.
%%The SDSS pipeline gave a redshift of 0.02943$\pm0.00007$.
The SDSS spectrum ($R\approx 1800$) was taken on 2000 December 21 UT with an exposure time of 5400\,s
 under the seeing of $\approx$2\arcsec, through a fiber aperture of 3\arcsec\ diameter.
As a fiber spectrum, the SDSS spectral data has not any spatial information available.
%% --- Note: that is the reason why we introduce the SDSS spectrum. -----

%% our MMT observation:
We conducted longslit spectroscopic observation on  2015 March~14 UT
with the Red Channel Spectrograph aboard the 6.5m MMT telescope.
 Two exposures of 900s each were obtained
 with the 1200 lines mm$^{-1}$ grating (blazed at 7700\AA)
 and a 1\arcsec-wide slit.
 The central wavelength is placed at 6800\AA.
  This provides a spectral resolution of $R\approx 3500$ (i.e., $\sigma_{\rm ins} \approx 36.5$\,\kms)
  and a wavelength coverage of 6410--7230\AA.
 The slit is oriented  at position angle (PA) 168$\arcdeg$ centered on \thisobj,
 with the nearby star SDSS~J083804.24$+$540619.9 also falling into the slit.
 The CCD was binned by a factor of two in the spatial direction, giving a spatial scale of 0\farcs6 per pixel.
 The blue-blocking filter L$-$42 was used to block light from higher orders.
 The seeing was 1\farcs2 and the sky was clear during the observation.

\subsection{Two-dimensional longslit spectrum}

Figure~\ref{fig:MMTimage} (panel~a) displays the MMT two-dimensional (2-D) longslit spectrum,
with the dispersion axis centered on the \ha\ line.
 The zero position of the spatial axis is set to be the spatial peak of the \ha\ emission.
 The most prominent feature is the extended \ha\ emission stretching out from the spatial zero position
 towards the two opposite directions.
%% The two \ha\ branches both extend to large spatial scales:
 The one towards the direction 12\arcdeg\ west of north (labeled as `N~branch' hereafter) extends 26 pixels
 (namely 15\farcs6 and a projected physical size 9.2~kpc),
 and the opposite one (labeled as `S~branch' hereafter) 29 pixels (namely 17\farcs4 and 10.2~kpc).
 %% 2016Sep19night ----
 Moreover, the spatial sizes of the two \ha\ branches are even larger than the circumgalactic ring (see panel~b).

\subsection{One-dimensional spectrum and spectral fitting}

The data reduction was performed with the standard routines in the IRAF program
to obtain the 1-dimensional spectrum.
\rev{%2017Jan24:---
Particular care is given to extracting the spectra of the two branches
that are extended and faint and even worse without a continuum;
e.g., we carefully select, by trial-and-error following IRAF manual,
the tracing parameters in the IRAF/apall task such as
\textit{t\_nsum}, \textit{t\_step} and the order of the trace fitting function.
}
We measured a redshift $z_{\rm em} = 0.02957\pm 0.00016$ from
the \oi$\lambda6300$ emission line that appears to have no spatially extended emission
and is susceptible little to possible star formation activity;
this redshift value is consistent with that given by the SDSS pipeline.
The rest frame of \thisobj\ hereinafter is referred to this redshift.

 Panels~c and d of Figure~\ref{fig:MMTimage}
 display the one-dimensional restframe spectra of the two branches of the spatially extended \ha\ emission.
 The extraction apertures are denoted in panel~a.
 The N~branch of the extended \ha\ emission is blueshifted by $\approx$162~\kms\
 as measured from the centroid of the best-fit \ha\ (namely the mean blueshift of the N~branch, see below and Table~1)
 relative to the theoretical wavelength in the rest frame,
 and the S~branch is redshifted by $\approx$86~\kms\ (namely the mean redshift of the S~branch).
 %%% 2017Jan27 about the velocity shifts above;  2017Jan24:
 \rev{We also extract the one-dimensional spectra of every lines of the spatial direction for the two branches,
 and then obtain the radial profile of the \ha\ flux per arcsec$^2$ (namely surface brightness)
 as illustrated in Figure~\ref{fig:ha_radial}.
 In the figure,  the gray shaded area denotes the region affected by \ha\ emission from the AGN BLR and NLR
 in the normal aperture; %%- 2017Jan27, figure caption.
 we do not extract the spectra for every spatial lines within the normal aperture because
 considerable starlight is present, and hard to model and subtract for every spatial line there.
 The radial surface brightness profile is irregular, not in a monotonic function with the projected radius $r$
 (e.g., by no means like $r^{-2}$).
 The lowest surface brightness of the extended \ha\ emission
 is about $6 \times 10^{-18}$ \flux\ arcsec$^{-2}$,
 which sets the spatial extent of the two branches  we can probe.
 In fact it is in terms of it that we set the outer bounds of the above extraction apertures for the two branches.}

 We need to verify that \thisobj\ has a real nuclear broad \ha\ component rather than
 a false appearance resulting from starlight contamination, or any velocity-offset \ha\ components
 emitted on larger scales (e.g., the two branches of the extended \ha\ emission).
 First, we extract one-dimensional spectra with two spatial apertures of different widths.
 One spectrum is extracted with a width of 5\farcs4 (namely 9~pixels and 3.2~kpc,
 hereinafter called ``normal aperture'';
 see in panel~a of Figure~\ref{fig:MMTimage} the aperture between the two red dash-dotted lines),
 which includes almost all light of \thisobj.
 The other spectrum is extracted with a width of 1\farcs2 only
 (hereinafter called ``small aperture'',
 equal to the seeing FWHM, corresponding to 2~pixels and 1.2~kpc;
 see in panel~a the aperture between the two blue dash-dotted lines)
 which only collects light from the central region of \thisobj.
 As panel~e shows, the two spectra of different extraction apertures
 have an almost identical shape of the \ha\ profile:
 the broad base of the \ha\ line profile---presumably the broad \ha\ line originated from the broad-line region (BLR)
 of the AGN---does not change with aperture size.
 This suggests that the broad \ha\ is not mimicked by narrow-line wings
 or any starlight features.
 Besides, we obtain a sum spectrum (see the dotted line in panel~e) by adding up
 the normal-aperture spectrum
 and the spectra of the two branches of the extended emission (see panels c and d).
 By comparing the sum spectrum with the normal-aperture spectrum,
 we find that
  the extended \ha\ emission does not impact the BLR-emitted \ha\ component at all,
  because (1) the extended \ha\ emission is relatively too weak,
  %% (particularly of the N~branch) %% -- removed, 2017Jan24 --
  and (2) more importantly
  the velocity offsets of the two branches are rather small
  (compared with the width of broad \ha, 2814~\kms\ FWHM; see panel~e),
  particularly of the S~branch that is the stronger branch
  but has a smaller velocity offset of $\approx$86~\kms\ only.
  %%% 2016Sep20afternoon -----

Next we perform continuum modeling and emission-line profile fitting.
We use the same fitting procedures as \citet{2012ApJ...755..167D}; here we only provide a brief description.
Regarding the MMT data, the normal-aperture (5\farcs4) spectrum is adopted,
corrected for Galactic reddening
using the extinction map of \citet{1998ApJ....500..525S} and the reddening curve of \citet{1999PASP..111...63F},
brought to the rest frame,
and then scaled to the (Galactic reddening corrected) SDSS spectrum%
\footnote{\,\thisobj\ is picked up from the IMBH sample of \citet{2012ApJ...755..167D} that was based on
the data set of the SDSS DR4.
Here we use the spectral data reduced by the improved SDSS pipeline, version {\it rerun 26},
as released since the SDSS DR7 (downloadable at http://das.sdss.org/spectro/ss\_tar\_26/\,),
instead of the data reductions archived in the SDSS DR4 (version {\it rerun 23}).}
according to the \oi$\lambda6300$ flux (see panel~a of Figure~\ref{fig:decompose}).

As the SDSS spectrum shows, the optical spectrum  is dominated by host-galaxy starlight.
According to the decomposition of \hst\ images (see \S\ref{sec:hstimage}), the AGN continuum is fairly weak
and can be safely ignored in the spectral fitting.
Because the wavelength coverage of the MMT spectrum is not sufficiently large to fit the starlight,
we obtain the best-fit starlight by fitting the SDSS spectrum, and then subtract it from the MMT one
to get the MMT emission-line spectrum.
The starlight is modeled with the templates of \citet{2006AJ....131..790L}, which were built from the spectra of simple
 stellar populations of \citet{2003MNRAS.344.1000B}.
 The starlight templates are broadened and shifted to match the stellar velocity dispersion of the galaxy, so that the stellar
 absorption lines are well subtracted. As shown in Figure~\ref{fig:decompose} (panel~a), the fit is good:
 the absorption features are well matched, and the residuals in the emission-line-free regions are consistent with the noise level.
%% 2016Nov23, after Feige's comment: --- %%
The strategy of subtracting the starlight from the MMT spectrum is reasonable
as the normal-aperture MMT spectrum comprises almost all the light (including starlight) in the slit from the source
and in fact matches the SDSS spectrum well (see panel a of Figure~\ref{fig:decompose}).

 Next, we fit the emission lines using the code described in detail in \citet{2005ApJ...620..629D}.
 The optical spectra show strong, narrow emission lines such as
 \oii, \oi$\lambda6300$, \ha, \nii$\lambda\lambda6548,6583$, and \sii$\lambda\lambda6716,6731$;
 \oiii$\lambda5007$ is instead weak.
 As we verified in the above, broad \ha\ is evidently presented \citep[see also][]{2007ApJ...670...92G,2012ApJ...755..167D}.
The SDSS data are used to fit the blue part such as \hb~$+$~\oiii\ region  (as not covered by the MMT spectrum),
  and the MMT data for the red part such as \oi\ lines, the \ha~$+$~\nii\ region and \sii\ lines.
 We fit the MMT spectrum first.
 Every narrow or broad components of the emission lines are modeled with Gaussians,
 starting with one Gaussian
 and adding in more if the fit can be improved significantly according to $F-$test.
 The \sii$\lambda\lambda6716,6731$ doublet lines are assumed to have the same profile
 and fixed in separation by their laboratory
 wavelength; the same is applied to \nii$\lambda\lambda6583,6548$ doublet lines,
 and to \oi$\lambda6300$ and \oi$\lambda6364$ emission lines.
 The flux ratio of the \nii\ doublet $\lambda6583/\lambda6548$ is fixed to the theoretical value of 2.96
 \citep[e.g.,][]{1989Msngr..58...44A,2000MNRAS.312..813S}.
 The best-fit model turns out that one Gaussian is used for the broad component of \ha,
 while two Gaussians for the narrow component of \ha,
 each line of the \nii, and \sii\ doublet and the two \oi\ lines, respectively.
 The reduced \chisq\ is 1.21 for the \ha~$+$~\nii\ region.
 %% Added after Paulina's comment, zoey : ---%%
Note that
 the \nii$\lambda6583$ and \sii\ doublet lines show asymmetric line profiles clearly in the MMT spectrum,
 which are actually common
 in high-resolution spectra \citep{2011ApJ...739...28X};
thus these narrow lines, and narrow \ha\ also, need two Gaussians to model.
 %%
 %%% --- then fit the blue part based on the SDSS data ---- %%%%
 Then we fit the \hb$+$\oiii\ region based on the SDSS spectrum.
 As the spectral signal-to-noise (S/N) of this region is not high and the \hb\
 emission line is much weaker than \ha,
 we assume the broad and narrow component of \hb\ have the same profiles
 as the respective components of \ha.
 The \oiii$\lambda\lambda4959,5007$ doublet lines are assumed to have the same profile,
 and the flux ratio of $\lambda5007/\lambda4959$ is fixed to the theoretical value of 2.98
 \citep[e.g.,][]{2000MNRAS.312..813S,2007MNRAS.374.1181D};
 the best-fit model turns out to adopt two Gaussians for each \oiii\ doublet line.
 The reduced \chisq\ is 0.95 for the \hb~$+$~\oiii\ region.
 \oii\ is fit independently, and one Gaussian is sufficient to model.
 %%
 %We also measure the extended \ha\ emission lines.
 %The line fluxes are directly integrated from the extracted spectra.
 We also fit the \ha\ lines of the N and S branches of the extended emission.
 One Gaussian is statistically sufficient for either branch, with reduced \chisq\ of
 0.95~(N) and 0.97~(S), respectively.
 The best-fit line parameters, centroid, FWHM (corrected for instrumental broadening),
 and flux of every branch agree well with those directly integrated/measured from the spectra.
 %% 2016Sep21, afternoon ----------

 All the best-fit line parameters are listed in Table~\ref{tab:linepar}.
 The best-fit model is illustrated in Figure~\ref{fig:decompose},
 for the \hb~$+$~\oiii\ region (panel~b) and the
 \ha~$+$~\nii\ region (panel~c).
 %% displayed in panels~b and c of Figure~\ref{fig:decompose},
 %% with a reduced \chisq\ of 1.06 for the \ha$+$\nii\ region and a reduced \chisq\ of 0.97 for the
 %%  \hb$+$\oiii\ region.
 We note that the broad-\ha\ FWHM is 2814$\pm$210~\kms, which
 is larger than the results of \citet[2120~\kms]{2007ApJ...670...92G}
 and \citet[2094~\kms]{2012ApJ...755..167D} based on the SDSS spectrum.
 This is probably because the MMT spectrum with a higher spectral resolution
 reveals the lower-contrast part of the wing of broad \ha\ than the SDSS,
 which can be discerned from the zoom-in view in the insert of panel~a of Figure~\ref{fig:decompose}.
 %%
 %% As the panel d in Figure \ref{fig:decompose}
 %%shows, the wing of \ha\ in the MMT spectrum is wider than that in the SDSS spectrum.

 \section{\hst\ imaging observations and analysis \label{sec:hstimage}}

  There are archival \hst\ images for \thisobj\
  observed by the WFC3 camera with the F160W (roughly the $H$ band of the Johnson system)
  and F555W (roughly $V$) filters (proposal ID: 12557, PI: K.~Gultekin),
  as well as the \hst/WFPC2 image with the F814W (roughly $I$) filter used by \citet{2011ApJ...742...68J}.
  The main body of the galaxy is almost round in shape and has no spiral arms,
  indicating an elliptical/spheroidal or a face-on S0 galaxy. However, a
  circumgalactic ring-like structure appears,
  prominent in the WFC3 images particularly (see Figure~\ref{fig:2ddecompose}).
  This ring is not very apparent in the WFPC2 image and
  was thus not mentioned by \citet{2011ApJ...742...68J}.%
  \footnote{\,Interestingly, we first noticed this ring from the SDSS image although the ring is faint
    and hard to discern there, and then we checked the HST archival data.}
%%%%
  We analyzed the WFPC2 image, and the result was almost the same as
  the structural fitting result presented by
  \citet{2011ApJ...742...68J};
  thus in the following we only describe our analysis of the WFC3 images.

  For ease of rejecting cosmic rays and to alleviate the
  undersampling problems of the point spread functions (PSFs),
  both the F160W and F555W observations were
  carried on in dither mode. The F160W image was targeted by the IR channel
  with a four-point dithering pattern called `WFC3-IR-DITHER-BOX-MIN' while the F555W
  image was taken with a three-step dither pattern named `WFC3-UVIS-DITHER-LINE-3PT'.
  The total exposure time for F160W and F555W band are 1073 and 1180 seconds, respectively.
  We begin the reduction with the flat-field calibrated images ({\it flt.fits})
  and combine the single exposures using standard software {\it Astrodrizzle}%
  \footnote{\,http://ssb.stsci.edu/doc/stsci\_python\_x/drizzlepac.doc/html/astrodrizzle.html},
  resulting in a final calibrated image for each band
  which is cosmic-ray-rejected,
  geometrically-distortion-corrected and sky-subtracted.

  Precise PSF is of great importance to separate the central AGN light from the host
  galaxy starlight \citep[e.g.,][]{2008ApJS..179..283K}.
  The WFC3/IR camera for the F160W image
  (with a plate scale of 0\farcs13 per pixel) is highly undersampled with PSF FWHM of
  $\approx0\farcs13$ (i.e., only $\approx1$~pixel).
  To achieve Nyquist sampling and also match with CANDLES hybrid PSF scale,
  we set {\it final\_scale}~=~0.06 to generate a new combined image with a higher
  sampling rate of 0\farcs06 per pixel. Besides, we set {\it final\_pixfrac}~=~0.8
  during the {\it Astrodrizzle} running in order to avoid introducing too much
  variation between pixels in the weight maps.
  The PSF FWHM of WFC3/UVIS camera for the F555W image
  (0\farcs0396 per pixel) is $\approx0\farcs07$,
  also slightly undersampled. We set {\it final\_scale}~=~0.03 in the {\it Astrodrizzle} combining.
  We adopted the hybrid PSF from CANDELs field, which works better than either modeled
  or empirical (stellar) PSFs \citep{2012ApJS..203...24V}%% 2017Jan24
  \footnote{\,They use 46 isolated stars in median-stacked form as an alternative PSF
    representation. The problem with this approach is the variety
    in the sub-pixel positioning of the stellar images, which leads to
    broadening in the PSF model when stacking a number of stars.
    In order to provide an accurate PSF model at all radii they take
    the median-stacked star and replace the central pixels (within
    a radius of 3 pixels from the center) by the TinyTim model
    PSF. The flux values of these pixels are normalized such that
    the total flux of the newly constructed hybrid PSF model is the
    same as that of the stacked star. Their hybrid PSFs have a uniform pixel scale of $0\farcs06$\,.}
   as the PSF of the F160W image.
  There is no existent hybrid PSF for the F555W image, so we turn to the modeled PSF
  produced by TinyTim\footnote{\,http://www.stsci.edu/hst/observatory/focus/TinyTim}.
  We first generate PSF to match with each single exposure, during which
  its position in the CCD and its defocus offset are carefully considered.
  Then we replace the central part of the object images by the corresponding
  PSF images and combine them using {\it Astrodrizzle} with the same drizzle parameters
  as above.

  Then we try to perform a 2-D decomposition of J0838 using GALFIT
  \citep{2002AJ....124..266P,2010AJ....139.2097P}.
  %(Peng~\etal\ 2002, 2010).
  The AGN is represented by a PSF, and the host galaxy is modeled by a bulge as a \citet{1968adga.book.....S}
  $r^{1/n}$ function or a disk as an exponential function
  (Exp; equivalent to $n = 1$ s\'{e}rsic)
  or a combination of them. The sky background has been subtracted during
  {\it Astrodrizzle} combination;
  the validity is confirmed by our residual sky estimation.
  The ring region (denoted by the two red ellipses in Figure~\ref{fig:2ddecompose})
  and other contaminations
  such as background galaxies or foreground stars are masked out.
  We take an fitting approach with incremental complexity of the model,
  which proves to be an effective way to obtain an optimal result
  \citep{2007ApJ...657..700D,2013ApJ...770....3J}.
%  (e.g., Dong et al. 2007; Jiang et al.  2013).
%%% ------- 2017Jan24, below:
%  We start the fitting with the simplest model, in which the host galaxy is represented
%  by a single s\'{e}rsic component. It yields obvious residuals with a large reduced \chisq.
%  %%
\rev{We start the fitting with the simplest model, in which the host galaxy is represented
  by a single s\'{e}rsic component in addition to a PSF for the AGN.
It yields obvious residuals with a large reduced \chisq.
Then we add an Exp disk component into the model,
and the fitting gets pretty acceptable for both bands.
  We try models with more free parameters (such as PSF plus two s\'{e}rsics),
  yet the fittings cannot be improved significantly.
  %% 2016Oct19
  Thus we adopt the \mbox{PSF$+$s\'{e}rsic$+$Exp} scheme as the best model.
 We note that,
 although it is quite weak (particularly in the F160W band)
 the PSF component is not dispensable, because
 (1) its existence is vindicated by the presence of the broad \ha\ emission line in the optical spectra (\S2.2)
 and (2) in practice without it the fittings would leave obvious residuals in the central region
  with large reduced \chisq\ particularly in the F555W image
  (see also Jiang \etal\ 2011 for the fitting of the F814W image). }
%% end of revision, 2017Jan24 ----.

%% To start a new paragraph, 2017Jan25:
  The fitting results are presented in Figure~\ref{fig:2ddecompose},
  and the best-fit parameters are summarized in Table~\ref{tab:galfit}.
%% below, Responding to the referee's comment:
\rev{%%2017Jan25:
In Figure~\ref{fig:2ddecompose} (bottom panels) we also plot the
1-dimensional representation of the data and the best fits, i.e.
the azimuthally averaged surface brightness
profiles of the \hst\ images and the GALFIT best-fit models as well as the residuals.
The surface brightness profiles are extracted with the IRAF/ellipse task that can interpolate into subpixel scale.
Note that
the interpolation into subpixel scale is somehow excessive and
the errors of the inner data points are inter-independent,
and thus those surface-brightness data are just for demonstration purpose and
used in Figure~\ref{fig:2ddecompose} only.
}

  The host galaxy is dominated by the bulge component,
  particularly in the F160W band.
  \rev{In the F160W band, the bulge has best-fit
  ellipticity $(1- b/a) = 0.29$ and s\'{e}rsic index $n=3.29$;  %% 2017Jan24
  the s\'{e}rsic index (around 4) is typical for merger-built classical bulges.}
  %% with an index of 3.29, a typical value for classical bulges.
  %%  --comment: n=1.83 in the F555W band
  Aside from the bulge, the disk is yet
  indispensable and can be easily fitted out from the 2-D images
  because the orientation of its major axis (PA $\approx44$\arcdeg) is almost perpendicular
  to that of the bulge component ($\approx-32$\arcdeg).
  While its best-fit size is similar in both filters (\mbox{$\approx1.0$\,kpc},
  see also the F814W result in \citealp{2011ApJ...742...68J}),
\rev{the disk becomes much prominent in the bluer band (namely F555W)  %%2017Jan24
with a best-fit ellipticity  $\approx0.1$,
  overshining the bulge outside $r=1$\arcsec\ (see Figure~\ref{fig:2ddecompose}, the left bottom panel).}
%
% // the text here and Figure~3  need to be revised by JN .....
%%% 2016Sep22, Thursday, noon. -----

%% to add a new paragraph, responding to the referee's comment; 2017Jan24 : ------
\rev{%%2017Jan24:
We would like to give more analysis and check about the use of our adopted hybrid PSF for the F160W image.
Potentially the PSF modeling impacts seriously the best-fit AGN and galactic disk components,
both of which (particularly the AGN) are much weaker with respect to the bulge component
in the NIR than in the optical (see Figure~\ref{fig:2ddecompose}).
The hybrid PSF is likely composed from substantially redder stars than the nucleus of \thisobj, and is thus wider.
This would create artificial flux transfer to some degree from the bulge component (namely the second most compact component)
to the AGN component in GALFIT fitting; certainly the inverse transfer is also conceptually possible
(particularly in sources with bright nuclei, which is not the situation of \thisobj).
To assess this possibility, we use TinyTim to generate a model PSF for the F160W filter.
We consider the CCD position, focus offset, AGN SED and so on,
following the aforementioned procedure for the F555W filter.
Then we run GALFIT again to fit the F160W image.
The fitting result changes little, with similar residual patterns.
%% the residual patterns:
Note that the spike-like white ``cross'' (namely over-subtraction to a certain degree)
in the residual image of the F160W fitting (Figure~\ref{fig:2ddecompose})
is not due to PSF mismatch, unlike the usual situation of GALFIT fitting images with bright AGN nuclei.
It is instead caused by the slight (as judged from the residual of the 1-dimensional representation
in Figure~\ref{fig:2ddecompose})
over-subtraction of the bulge and disk components;
as mentioned in the above the two are oriented almost perpendicularly.
Besides, the best-fit disk in the F160W is by no means only to compensate possible PSF mismatch,
because (1)
the disk component is indispensable as we argue in the above paragraph with the F555 fitting result
(as well as the F814W of Jiang~\etal\ 2011),
(2) the AGN component is rather faint in the F160W band and cannot leave considerable residual fluxes
on the scale of the galactic disk,
and (3) the best-fit disk parameters (size, PA and ellipticity) of the F160W image
are consistent with those of the optical images.
%%%
We have also extracted the 1-dimensional surface brightness profiles of both the hybrid PSF
and the TinyTim F160W one, and find that the discrepancy between the two is almost negligible at all radii.
In fact, Kim~\etal\ (2008) have conducted detailed tests (see their \S2.1)
and concluded that for HST PSFs
the variation due to SED is \mbox{$<\,10$\%} at all radii,
which is negligible compared with that due to spatial distortion, temporal variability and particularly pixel undersampling.
Certainly, as demonstrated by the 1-dimensional surface brightness profiles in Figure~\ref{fig:2ddecompose} (bottom panels)
the bulge component overwhelms the other two components (the AGN and the disk) in the NIR,
and hence we caution that the best-fit parameters for the AGN and the disk from the F160W image  are not as reliable
as from the F555W image.
It is for this very reason that in the above we quote the information (s\'{e}rsic index, size, PA, and ellipticity)
for the bulge from the F160W fitting
and for the galactic disk from the F555W fitting.
}%% end rev %%2017Jan24

%  \section{Results and Discussion}
 %% Section 4
 \section{The IMBH and LINER 1 nucleus}
    %\subsection{The AGN Properties and the Host Galaxy}
  %nucleus

     Based on the analyses of the spectra and images, we are able to estimate the properties of the host
     galaxies and central BH for \thisobj.
%%     The optical and NIR \hst\ images are dominated by a smooth, almost round in shape
%%     structure.  %%-- noting to do with the AGN!
     The 2-D decomposition of the \hst\ images by this work and \citet{2011ApJ...742...68J} show that \thisobj\ incorporates
     three components: a bright bulge component, a weak galactic disk component, and a faint PSF component.
     Spectral analysis reveals that \thisobj\ has a weak but evident broad \ha\ emission line originated from the AGN BLR.
     %%%
     This object was covered by the VLA FIRST survey \citep{1995ApJ...450..559B}.
     We analyze the radio data and find a compact source
     located in the nucleus position with a considerably high
     statistical significance, and the measured flux at 20cm is 1.09~mJy with a rms noise of 0.15~mJy.%%
   \footnote{\,We use the program {\it Gaia} in the starlink software package (http://starlink.eao.hawaii.edu/starlink/\,)
     to fit a 2-dimensional Gaussian to the source at the position of \thisobj\ in the VLA FIRST image.
     The source turns out to be point-like, with a flux of 0.84~mJy and rms
     noise of 0.15~mJy.
     The source detection threshold set by the FIRST catalog is 1~mJy, so this source
     was not included in the official catalog.
     With a CLEAN bias (which always makes an underestimate of the flux)
     of 0.25~mJy \citep{1995ApJ...450..559B}, the real flux of the source should be 1.09 mJy,
     7.3 times the rms noise.}
     We calculate the k-corrected radio power as $P_{\rm 20cm} = 4 \pi D_L^2 f_{\rm FIRST} /(1+z)^{1+\alpha_r} $,
     where the radio spectral index $\alpha_r$ ($F_\nu \propto \nu^{\alpha_r}$) is commonly assumed to be $-0.5$.
     %% With $f_{\rm FIRST} = 1.09$~mJy,
     Then we get $P_{\rm 20cm} = 2.16\times10^{21}$ W~Hz$^{-1}$.
%% radio loudness --- zoey :
     We calculate the radio-to-optical flux ratio (namely radio loudness) following \citet{2002AJ....124.2364I},
     $R_{i} \equiv \log(\frac{f_{20\rm cm}}{f_{i}})$, where $f_{i}$ and $f_{20 \rm cm}$ are flux densities ({\mbox{per Hz})
     at the $i$ band and 20~cm, respectively.
     %% The $i$-band flux density of the AGN  is 0.011~mJy (Galactic-reddening corrected),
     The thus-calculated $R_{i}$ is 2.0, indicating that \thisobj\ is a mildly radio-loud AGN.
%%
       %%%%% 2016Sep22 afternoon: -----%%%%
     %%All of these above indicate that \thisobj\ has nuclear activities, though it is not very high.

%% to start a new paragraph: -- 2016Nov23:  --- %%
     With the measured luminosity and line width of the broad \ha\ emission line,
     the BH mass can be estimated with
     the commonly used virial mass formalisms that adopt the empirical relation between BLR radius and AGN luminosity
     (the $R$--$L$ relation; \citealt{2000ApJ...533..631K}).
     The BH mass values given by \citet{2007ApJ...670...92G} and \citet{2012ApJ...755..167D} are 1.58 $\times 10^{6}$ \msun\ and
     1.26 $\times 10^{6}$ \msun\ respectively.
     Here we estimate the BH mass with our measurement of the \ha\ emission line from the MMT spectrum (see \S2.2),
     %% \footnote{\,The flux level of
     %% the MMT spectrum is scaled to the SDSS spectrum.}
     using the mass formalism given by \citet[their \mbox{Eq. 6}]{2011ApJ...739...28X},
%     \begin{equation}
%       \log(\frac{M_{\rm BH}}{\msun})\ =\ 6.40^{+0.09}_{-0.07}\ +\ (0.45\pm0.05)\,\log(\frac{L_{\rm H\alpha}}{10^{42}\, \rm \ergs})
%       \ +\ (2.06\pm0.06)\,\log(\frac{\rm FWHM(broad~H\alpha)}{10^3\,\kms}) \ \ .
%     \end{equation}
    % This formalism
     which is based on \citet{2005ApJ...627..721G,2005ApJ...630..122G}
     but incorporates the $R$--$L$ relation recently calibrated by \citet{2009ApJ...697..160B}.
     The total (BLR $+$ NLR) \ha\ luminosity measured from the normal-aperture MMT spectrum
     is \rev{$L_{\rm \ha} = 1.45\times10^{40}$ \ergs}.
     Note that according to the broad-line \ha/\hb\ ratio (3.4; see Table~1),
     the broad lines suffer nearly no dust obscuration (see \citealt{2008MNRAS.383..581D}).
     %% 2016Nov19 . %%
     %%Together with FWHM(\ha) = 3109\,\kms,
     \rev{Thus we obtain $M_{\rm BH} = 3.15 \times 10^{6}$ \msun }.
     This value is larger by a factor of \rev{ $\approx$ 2}
     than those given by \citet{2007ApJ...670...92G} and \citet{2012ApJ...755..167D},
     mainly because of our larger FWHM of broad \ha\ measured from the MMT spectrum;
     this discrepancy is roughly within the uncertainty of virial mass estimation
     \citep[e.g.,][]{2009ApJ...707.1334W,2011ApJ...739...28X}.

     To estimate the bolometric luminosity ($L_{\rm bol}$), we follow the recipe of \citet{2007ApJ...670...92G} for
     the bolometric correction from $L_{\ha}$,
     %%according to $L_{\rm bol}=2.34\times 10^{44}(L_{\ha}/10^{42})^{0.86}$.
     yielding \rev{$L_{\rm bol} = 6.21 \times 10^{42}$~\ergs}.
     The corresponding Eddington ratio is thus \rev{$L_{\rm bol}/L_{\rm Edd} = 0.016$}.
    %% 2016Nov23---: %%
     The $L_{\ha}$ of \thisobj\ (and correspondingly the $L_{\rm bol}$)
     is slightly lower than the median $L_{\rm \ha}$ ($1.9\times10^{40}$ \ergs)
     of the nearby type-1 Seyfert galaxies and quasars listed in Table~1 of \citet{2008ARA&A..46..475H},
     but higher than the median
     of the nearby type-1 LINERs listed there ($3.7 \times 10^{39}$ \ergs).
     The broad-\ha\ luminosity is slightly larger than that of
     the IMBH AGN in a less luminous galaxy at a similar redshift
     (SDSS~J160531.84$+$174826.1, $z=0.032$; \citealt{2007ApJ...657..700D}),
     but lower than that of the prototypal IMBH AGN, POX~52 at $z=0.022$ (see Table~4 of \citealt{2007ApJ...657..700D}).

% The monochromatic continuum luminosity $\lambda L_{\lambda}$(5100 \AA), which can be estimated from the total \ha\
% luminosity\citep[their Eq. 1]{2005ApJ...630..122G}, is $\lambda L_{\lambda}$(5100 \AA) = 2.76 $\times 10^{41}$\ergs.
% Using the conversion $L_{\rm bol} \approx 0.75\times L_{\rm iso}$ where
% log$L_{\rm iso}$=4.89$+$0.91log$\lambda L_{\lambda}$(5100\AA) \citep{2012MNRAS.422..478R}, the bolometric luminosity is
% estimated to be $L_{\rm bol} = 6.17 \times 10^{42}$~\ergs, and the corresponding Eddington ratio is
% $L_{\rm bol}/L_{\rm Edd} = 0.013$.

     %% new paragraph here:  2016Sep22 afternoon ------

     \thisobj\ is among the ten LINER~1 galaxies in the \citet{2012ApJ...755..167D} sample of 309 low-$z$ IMBH AGNs.
    \rev{ From the SDSS and MMT spectra (see Table~\ref{tab:linepar}), the optical narrow
     emission lines are observed to have the following line-intensity ratios:
     \mbox{log\,\oiii$\lambda5007$/\hb} = \mbox{$-0.09\pm0.11$},
     \mbox{log\,\nii$\lambda6583$/\ha} = \mbox{$-0.81\pm0.12$},
     \mbox{log\,\sii$\lambda\lambda6716,673$/\ha} = \mbox{$-0.28\pm0.04$},
     and \mbox{log\,\oi$\lambda6300$/\ha} = \mbox{$-0.52\pm0.05$}.}
     In the BPT diagnostic diagrams \citep{1981PASP...93....5B,2003MNRAS.346.1055K,2006MNRAS.372..961K},
     \thisobj\ corresponds to the boundary case between \hii\ and LINER:
     it is located at the \hii\ portion
     in the \oiii/\hb\ versus \nii/\ha\ diagram,
     on the border between \hii\ and LINER in the \oiii/\hb\ versus \sii/\ha\ diagram,
     and at the LINER portion in the \oiii/\hb\ versus \oi/\ha\ diagram.
     We also note its strong \oii\ and \oi$\lambda6300$ emission lines, which are characteristic of LINERs.
     \rev{With the flux ratios \mbox{log\,\oii/\oiii$\lambda5007$} = \mbox{0.46$\pm$0.11} and
     \mbox{log\,\oi$\lambda6300$/\oiii$\lambda5007$} = \mbox{0.17$\pm$0.11}},
     \thisobj\ safely satisfies the criteria of being a LINER of
     Osterbrock \& Ferland (2006; \oii/\oiii$\lambda5007$ \mbox{> 1} and \oi$\lambda6300$/\oiii$\lambda5007$ \mbox{> 1/3}).
     %% 2016Nov23 .
     %%\citet[\oii/\oiii$\lambda5007 \mbox{ > 1}$ and \oi$\lambda6300$/\oiii$\lambda5007 \mbox{ > 1/3}$]{2006agna.book.....O}.
     %%
%     Osterbrock~\& Ferland (2006;
%     %%\citep{2006agna.book.....O}
%     \oii/\oiii$\lambda5007 >~1$ and \oi$\lambda6300$/\oiii$\lambda5007 >~1/3$).
%    To sum up, \thisobj\ can be classified as a LINER,
%    of the broad-line version (namely LINER~1; \citealt{1997ApJ...487..568H}).

 %% Section 5
 \section{The host galaxy, ring and extended \ha\ emission}
     %$M_{\rm V}$ = -17.93 mag

    \thisobj\ is an early type galaxy of a red color with $u-r = 2.41$, which satisfies the red galaxies criteria
    ($u-r \gtrsim 2.22$) of \citet{2001AJ....122.1861S}.
     With a total luminosity (bulge $+$ disk $+$ ring) of $M_{V} = -17.80$~mag measured from the \hst\  F555W image,
     it is a dwarf galaxy, even fainter than the Large Magellanic Cloud (LMC, $M_{V} = -18.5$~mag).
     The SDSS spectrum is dominated by starlight from old stars. The stellar age of a galaxy can be roughly estimated from
     the 4000-\AA\ break index $D(4000)$ and the \hd\ absorption index (H$\delta_{A}$); see \citet{2003MNRAS.341...33K}.
     We measure $D(4000)$ and
     H$\delta_{A}$ from the SDSS spectrum and obtain $D(4000)$ = 1.66, H$\delta_{A}$ = 0.40. According to
     Figure~2 of \citet{2003MNRAS.341...33K}, both the indices indicate the mean stellar age of \thisobj\ is about 2.5~Gyr.
     We use the K-band luminosity of the starlight to calculate the stellar mass, because the mass-to-light ratio in the
     K band ($M/L_{\rm K}$) is relatively insensitive either to dust absorption or to stellar population age.
     \citet{2013MNRAS.430.2715I} provide a calibration of mass-to-light ratios against galaxy colors; here we
     use  \mbox{log\,$M/L_{\rm Ks} =$} $0.794(g - i) \,-0.997$ (with a scatter of $\pm$0.1 dex; see their Table~3).
     The $g$ and $i$ are the SDSS Petrosian magnitudes
     and $L_{\rm Ks}$ is calculated from the 2MASS
     $K$ band of \thisobj, giving
     $g - i = 1.19$ and $L_{\rm Ks} = 5.00 \times 10^{9}\,L_\sun$. %5.02 \times 10^{9}$\lsun.
     Then the stellar mass is $M_{\rm host} = 4.45 \times 10^{9}$\msun.
    %The color of \thisobj\ is red, and its optical spectrum is dominated by old stellar components. These suggest it is
     %probably a gas poor galaxy and with low starformation rate (SFR).

     According to \oii\ emission line from the SDSS spectrum, we can estimate the star formation rate (SFR)
     of the nuclear region ($r \lesssim 0.9$~kpc).%%
     \footnote{\,The SDSS spectrum is extracted through a fiber aperture of 3\arcsec\ in
     diameter, i.e., a radius (projected physical distance) of 0.9~kpc at this source's redshift.}
     But \oii\ is a low-ionization forbidden line and can be excited in either \hii\ regions,
     or in narrow line regions (NLRs) of AGNs,
     or in a shock-heated plasma.
     It is hard to estimate the fraction of \oii\ coming from possible \hii\ regions for \thisobj.
     If we assume {\em all} the \oii\ flux is from \hii\ regions of \thisobj,
     we can use the extinction-corrected \oii\ luminosity to get the upper bound to the SFR.
%%     The internal extinction of narrow lines is determined by comparing the observed Balmer decrement \ha/\hb\ of narrow
%%     component to the theoretical value for case~B recombination.
%% NLR is not in Case B. -- 2016Nov20 . %%
     The extinction of narrow lines can bee determined
     by comparing the observed narrow-line Balmer decrement \ha/\hb\
     with the intrinsic value in AGN NLRs (3.1; \citealt{1983ApJ...269L..37H,1984PASP...96..393G}).
     \rev{Assuming the extinction curve to be the Small Magellanic Cloud (SMC) one \citep{2003ApJ...594..279G},
     we get $E(B-V) = 0.38$. The measured
     $L_{\rm [OII]\lambda3727}$ is $4.36\times10^{40}$~\ergs,
     and the extinction-corrected $L_{\rm [OII]\lambda3727}$ is
     $2.13\times10^{40}$~\ergs.
     Employing the calibration of \citet[their Eq.~10]{2004AJ....127.2002K}, we derive the upper bound to SFR(\oii)
     to be \mbox{0.15 \msun\ yr$^{-1}$}}.

  %% the ring ----
     \hst\ images reveal a circumgalactic ring structure of \thisobj.
     The ring is an ellipse in the sky, with its major and minor axes being 27\arcsec\ (15.6~kpc)
     and 19\arcsec\ (11.4~kpc),
     respectively, in the F160W ($H$) band.
     It exhibits a disperse shape on its south side in the sky, which spreads
     $\sim 6\farcs5$ along the radial direction.

%% the rotation-like Halpha emission ---
     The most striking feature of \thisobj\ is the large-scale \ha\ emission extending to 10~kpc from the center
     of the galaxy in the MMT 2-D longslit spectrum.
   %% 2016Nov23: -- %%
    We measure the distributions of the line-of-sight (LOS) velocity of \ha-emitting gas (namely \ha\ line profiles) at
    different spatial positions along the slit.
  %% 2017Jan27:
  To increase the S/N  (compared with Figure~\ref{fig:ha_radial}),
  we combine the spatial lines every 2 pixels.
Velocity measurements are also done for the spatial lines within the normal aperture (namely the main body of the galaxy)
since such a kind of measurement does not require exact flux calibration and measurement;
the normal apertures (9 pixels) are divided evenly into 3 bins.
    The distributions are close to normal, and we then fit them with a Gaussian each.
    From the fitting results we obtain
    the mean velocity ($v$, namely the centroid of the Gaussian distribution, with respect to the systematic redshift) and
     velocity dispersion ($\sigma$) at every spatial positions;
     the 1-$\sigma$ errors are also given by the fitting.
     In Figure~\ref{fig:rotationcurve} we plot
     the radial profiles of $v$ (top panel)
     and $\sigma$ (bottom panel).
     In the spatial axis $r$, zero denotes the position of the central AGN in the slit (see Figure~\ref{fig:MMTimage})
     and north is toward positive values.
     The positive $v$ values denote red shifts, and the negative, blue shifts.
     The $\sigma$ values are not corrected for instrumental broadening. %%2016Nov23
%% below 2016Nov24 --- : %%
\rev{%% 2017Jan27:  about Figure 5 ---:
The $v$ profile shows a
     steep decrease by \mbox{$\approx$250\,\kms}
     when $r$ runs from $-4\arcsec$ (\mbox{2.4\,kpc} south) to $+4\arcsec$ (\mbox{2.4\,kpc} north).
     With $r$ going outwards along the S branch
     $v$ basically keeps constant at the mean value as measured from the integrated S-branch aperture (86\,\kms),
     marginally reaching the maximum redshift ($130\pm46$\,\kms) at $r \approx 5$\,kpc south
     and then dropping and roughly leveling around the mean.
     On the positive $r$ part (namely the N branch),
     $v$ basically keeps constant at the mean value as measured from the integrated N-branch aperture (162\,\kms),
     marginally reaching the maximum blueshift ($200\pm46$\,\kms) at $r \approx 5.4$\,kpc north
     and then decreasing again and roughly leveling at $\sim$140\,\kms\ towards the  larger north end ($r \gtrsim 7.5$\,kpc).
    %%%
     The $\sigma$ profile, except in the central region (roughly $r$ in the range [$-2$, 2]\,kpc)
     that is affected by the \ha\ emission from the AGN BLR and NLR,
     are basically flat with velocity dispersion values in the range \mbox{$\approx30-90$\,\kms}
     (without correcting for instrumental broadening).
     Considering the instrumental broadening $\sigma_{\rm ins} = 36.5$\,\kms,
     the intrinsic velocity dispersion of the extended \ha-emitting gas is very small;
     only in the two positions with the maximum redshift or blueshift
     (namely $r$ around 5\,kpc south and 5.4\,kpc north, respectively)
     $\sigma$ reaches a maximum $\approx80$\,\kms\ (corrected for instrumental broadening).
}%% end of 2017Jan27.
     Both the $v$ and $\sigma$ profiles are similar to many rotation disks of warm gas reported in the literature
     \citep[e.g.,][]{2010ApJ...721..762W,2011ApJ...740...83K}.

    %% starting a new paragraph -- 2016Nov25: ---
     In addition, we do not detect significant starlight continuum on the large scales of
     the extended \ha\ emission, neither in the \hst\ images
     nor in the MMT longslit data;
     this is evident in Figure~\ref{fig:2ddecompose}
     and indicated by the continuum levels (close to 0) in panels c and d of Figure~\ref{fig:MMTimage}.
     It is also worth noting that the physical scale of the extended \ha\ emission
     corresponds to
     or even larger than
     the faint outer rim of the ring as revealed by the \hst\ images (Figure~\ref{fig:2ddecompose}).
     Several bright spots of the extended \ha\ emission along the slit (e.g., at $r = 6$\,kpc south)
     roughly correspond to the inner rim of the ring.

 %% Section 6
     \section{{\it In situ} star formation powers the extended \ha}

     %The most striking feature of \thisobj\ is the large scale extended \ha\ emission extending to 10~kpc from the center
     %of the galaxy in the MMT 2-D longslit spectrum.
     The presence of the spatially extended \ha\ emission (see \S2.1) indicates that
     ionized warm gas exists on the scales 10~kpc away from the
     center of the galaxy.
     In the literature such large-scale emission lines were often reported
     in dwarf starburst galaxies as powered by
     {\it in situ} star formation activities \citep[e.g., NGC~1569,][]{2002ApJ...574..663M},  %% -- 2016Nov19 -- %%
     but seldom in galaxies with such red colors.
     Thus two questions emerges naturally as follows.
     The first question --- what powers the large-scale \ha\ emission line for this object: AGN, star formation, or
     heating by some kind of shock?
     The second question --- where is the ionized gas from?

%% 2016Nov19 night --:  %%%
     As to the first question,
%% 2017Jan25:
 assuming all the \ha\ flux is from hydrogen recombination,
 the total number of ionizing photons
     must be large enough to balance
     the total number of recombination events in the ionized gas.
Following
     Eq.~(13.4) of \citet{2006agna.book.....O}, we can write the luminosity of \ha\ as follows,
     \begin{equation}
     \label{eq:Lha_Q}
       L_{\ha} = hv_{\ha}\frac{\alpha_{\ha}^{eff}({\rm H}^{0}, T)}{\alpha_{B}({\rm H}^{0},T)}\frac{\Omega}{4\pi}Q_{\rm H^{0}}
       \approx 1.37\times10^{-12}\frac{\Omega}{4\pi}Q_{\rm H^{0}}  ~~ ,
     \end{equation}
     where $\frac{\Omega}{4\pi}$ is the nebular covering factor, $\frac{\alpha_{\ha}^{eff}({\rm H}^{0}, T)}{\alpha_{B}({\rm H}^{0},T)}$
     is the number of \ha\ photons produced per hydrogen recombination,%% -2016Nov19 -- %%
     %% footnote for calculation: --- %%
     \footnote{\,The parameter $\alpha_{\ha}^{eff}({\rm H}^{0}, T)$
     is the effective recombination coefficient that reflects the number of recombinations
     to all $n\geq3$ levels which ultimately lead to transitions from $n=3$ to $n=2$ (namely emitting \ha\ photons).
     The parameter $\alpha_{B}({\rm H}^{0},T)$ is the Case~B recombination coefficient.
     At \mbox{$T_{\rm e} = 10^{4}$\,K} and under Case~B condition (see Tables 2.1 \& 4.2 of \citealt{2006agna.book.....O}),
     $\alpha_{B}({\rm H}^{0},T)$ is $2.59\times10^{-13}$ \mbox{cm$^{3}$\,s$^{-1}$},
     $\alpha_{\hb}^{eff}({\rm H}^{0}, T)$ is $3.03\times10^{-14}$ \mbox{cm$^{3}$\,s$^{-1}$},
     and the Balmer line intensity ratio %% 2016Nov20  %%
     $j_{\ha}/j_{\hb} = 2.87$.
     According to
     $\frac{\alpha_{\ha}^{eff}({\rm H}^{0}, T)}{\alpha_{\hb}^{eff}({\rm H}^{0}, T)} = \frac{j_{\ha}}{j_{\hb}}\frac{\lambda_{\ha}}{\lambda_{\hb}}$
     \citep[cf. Eq.~4.14 of][]{2006agna.book.....O},
     $\alpha_{\ha}^{eff}({\rm H}^{0}, T)$ is $1.17\times10^{-13}$ \mbox{cm$^{3}$\,s$^{-1}$}.}
     %%% end of the footnote --- %%%
     and $Q_{\rm H^{0}}$ is the number of photons that can ionize hydrogen atoms.
     %%% 2017Jan25:
     By means of Eq.~(\ref{eq:Lha_Q}), then we can estimate the ionizing photons required to produce the observed \ha\ luminosities
     of the broad component, the narrow component and the extended,
     provided that the corresponding covering factors ($\frac{\Omega_{\rm b}}{4\pi}$, $\frac{\Omega_{\rm n}}{4\pi}$ and $\frac{\Omega_{\rm e}}{4\pi}$) are known.
     The measured broad, narrow and extended \ha\ luminosities are \rev{$5.73\times10^{39}$\ergs,
     $8.78\times10^{39}$\ergs, and $1.22\times10^{39}$\ergs }, respectively.

     %% to start a new paragraph: 2017Jan25: ---
     We first consider the possibility that the ionizing photons are all provided by the central AGN.
     Adopting the ionizing continuum of the median quasar SED
     (either radio-loud or radio-quiet) of \citet{1994ApJS...95....1E} and
     scaling it to the observed $\lambda L_{\lambda}$(5100\AA) of this object,
     we obtain the number of the ionizing photons
     $Q_{\rm H^{0}} = \int_{\rm 13.6\,eV}^{\infty} \frac{L_{\nu}}{h\nu} d\nu = 1.5\times10^{52}$ per second.
     We note that previous studies showed that
     %% the SED of LINERs are likely short of the big blue bump %% 2017Jan24
     the SED of LINERs does not likely have the big blue bump
     %% the big blue bump in the LINERs SEDs is weak or even absent
          \citep[e.g.,][]{2008ARA&A..46..475H,2012A&A...539A.104Y}
     and thus the number of the ionizing photons  of \thisobj\ (being a LINER)
     should be less than the above  $Q_{\rm H^{0}}$ value.
%%%%%
     On the other hand,
     typically in AGNs
     the BLR covering factor $\frac{\Omega_{\rm b}}{4\pi} \sim 0.1$
     and the NLR one $\frac{\Omega_{\rm n}}{4\pi} \sim 0.07$ \citep{2009ApJ...705..298M}.
     Considering its disk configuration (\S5),
     the covering factor of the extended \ha-emitting gas
     to the central source should be very small (say, $\frac{\Omega_{\rm e}}{4\pi} \ll 0.01$).
     According to Eq.~(\ref{eq:Lha_Q}),  %%2017Jan25
     the required number of the ionizing photons
     accounting for the broad \ha\ luminosity is $4.18 \times 10^{52}$ per second;
     $9.16 \times 10^{52}$ per second for the narrow \ha;
     $8.96 \times 10^{50}/\frac{\Omega_{\rm e}}{4\pi}$ per second for the extended \ha.
     %% the total \ha\ luminosity (broad $+$ narrow $+$ extended) should be
     %% \rev{$1.33\times 10^{53}+ 8.91\times 10^{50}/\frac{\Omega_{\rm e}}{4\pi}$ per second}. %zoey, 2017Jan
     %%% 2016Nov20: --- %%%
     Note that here we do not count emission lines other than \ha,
     and thus the real required ionizing photons should be much more than the above estimated.
%%%%%
Hence we can see that the AGN-provided ionizing photons may roughly balance
the one required by the broad \ha\ emission line only,
but is not able to further power
all the narrow \ha\ emission line extracted from the normal aperture,
let alone the large-scale \ha\ that should have a tiny covering factor to the central AGN.
%%%
%%% Wenjuan added, after the comment by Paulina : --- %%
     There might be a possibility that the warm extended gas is powered by stronger AGN activity in the past time,
     about $3 \times 10^5$ years ago (corresponding to the distance of the extended \ha\ emission to the AGN). %% 2016Nov19 -- %%
     This is however disfavored by the line-ratio diagnostics of the extended emission (see below).

 %%%%%%%%%%
%% 2017Jan25, now post-AGB:
\rev{%% 2017Jan25, now the case of post-AGB:
In light of the old stellar age of the host galaxy,
we then consider hot evolved stars as a probable energy source.
In galaxies with stellar populations older than \mbox{$\sim1$ Gyr} ,
evolved stars after their asymptotic giant branch phase (post-AGB)
can become sufficiently hot ($\sim 10^{5}$~K) to produce a radiation field
capable of ionizing atoms with a substantial
ionization potential \citep{1994A&A...292...13B,2008ARA&A..46..475H,2013A&A...558A..43S}.
In terms of the BPT diagrams, the emission lines powered by post-AGB stars
appear like a LINER spectrum that is just the case of the narrow emission lines of \thisobj.
Post-AGB stars have a specific ionization rate of $Q_{\rm ion} = 7.3 \times 10^{40} \frac{M}{\msun}$~s$^{-1}$
 \citep{1994A&A...292...13B,2008ARA&A..46..475H}.
Multiplied with our derived stellar mass of \thisobj\ ($4.45 \times 10^{9}$\msun),
the post-AGB stars embedded in the whole galaxy
can contribute $3.25 \times 10^{50}$ ionizing photons per second.
This value is lower by 1 order of magnitude
than either the required by the broad \ha\ emission even if assuming a complete covering ($\frac{\Omega_{\rm b}}{4\pi} = 1$)
 or the required by the narrow \ha\ emission assuming $\frac{\Omega_{\rm n}}{4\pi} = 1$,
 or is 2.7 times lower than the required by the extended \ha\ emission assuming $\frac{\Omega_{\rm e}}{4\pi} = 1$.
 It can be regarded reasonable to assume a complete covering of the narrow-\ha-emitting gas to the post-AGB stars
 because the two are generally co-spatial.
It is however not possible for the extended \ha-emitting gas  to completely cover the post-AGB stars,
because the post-AGB stars are predominantly distributed within the main body of the galaxy (namely within $r \approx 3\arcsec$)
but the extended \ha-emitting gas is much farther out and  likely in a thin-disk configuration (\S5).
It thus appears plausible that, considering possible uncertainties in the above estimations,
the AGN continuum powers the broad \ha\ emission line,
the post-AGB stars may power (part of) the
     narrow \ha\ emission line,
but neither is able to further power the large-scale \ha\ emission.
This inference is consistent with the radial surface brightness profile of the extended \ha\ emission that is irregular
without a monotonic dependence with projected radius (\S2.2).
Actually it is vindicated  by the \ha-involving BPT diagram of the normal-aperture narrow lines (\S4)
     and the BPT diagnostics of the extended emission below.
} %% end of the post-AGB case, 2017Jan25.

     %% to start a new paragraph -- 2016Sep22night ---
     We do not detect any significant spatially extended emission features of low-ionization lines (\oi, \nii, and \sii) in the
     2-dimensional longslit spectrum.
%   %
%    %%%% 2016Oct7 revised:
%    %% However, this cloud be due to the weakness of these lines or because of the contamination
%    %% from the nearby sky lines (e.g., \sii\ doublet, \oi).
%      %We note that low-ionization lines (\oi, \nii, and \sii) do not have such obvious large scale extended emission features in
%      %the 2-D longslit spectrum.
     The \oi$\lambda6300$ line is located at the blue end of the MMT spectrum where the spectral quality is poor.
     The \sii\ doublet lines suffer from contamination from the nearby sky lines.
     For the extended \nii,
     we try to measure the upper bound of the flux from
     %% 2017Jan27
     %% the spectrum of the S branch that has a higher S/N than the N branch.
     the spectra of both branches.
     As the \nii\ doublet lines in either branch
     are rather weak (if not absent), we calculate the flux and flux error for the \nii$\lambda6583$ line
     directly from the spectrum of each branch (see panels~c and d of Figure~1),
     which gives \rev{$7.1\pm2.4 \times 10^{-17}$~\flux\ for the N branch
     and $4.6\pm 2.9 \times 10^{-17}$~\flux\ for the S branch}.
     We take the thus-estimated flux plus 2 times the flux error
     as the upper bound to the \nii$\lambda6583$ flux,
     i.e. \rev{$1.19 \times 10^{-16}$~\flux\ for the N branch and $1.04 \times 10^{-16}$~\flux\ for the S branch}.
 %%%%
     We confirm that these values are definitely a secure upper limit, by plotting the \nii$\lambda6583$ line
     assuming the line having the same width as the corresponding \ha\ and the above upper-limit flux for each branch.
     \rev{Thus the line ratio \mbox{log\,\nii$\lambda6583/\ha$} \mbox{$< -0.38$} for the N branch
     and \mbox{log\,\nii$\lambda6583/\ha$} \mbox{$< -0.50$} for the S branch,
     both being in the \hii\ region of the BPT diagram.}
     This further supports our inference that the extended, line-emitting gas is powered by star burst.

     %% 2016Oct7 :
     It is also unlikely that the large-scale \ha\ line is powered by shock heating,
     at least not primarily.
     As model calculations suggested in the literature,
     shocks usually enhance those low ionization forbidden lines \citep{2008ApJS..178...20A},
     which is opposite to the situation in \thisobj.
     %% However,
     Certainly, conceptually there are still several other possibilities accounting for the absence of
     low-ionization emission lines on the large spatial scales,
     e.g., the suppression for low-ionization lines due to high ionization state,
     the low-metallicity of the gas, etc.
     But, low-metallicity gas is not common in the local Universe; the AGN of
     \thisobj\ is neither in a high ionization state nor seriously obscured (\S4),
     unlike the AGN-powered cases in forms of either direct photoionization or shock heating
     \citep[e.g.,][]{2009ApJ...690..953F,2016arXiv160904021C}. %%2017Jan24 .

%% to start a new paragraph: 2016Nov19---: %%
     Summing up the above,
     we therefore conclude that the ionizing photons for the large-scale emission
     are mainly provided by the {\it in situ} star formation.
     According to the empirical relation
     %% $SFR /\msun \rm yr^{-1} = 7.9 \times 10^{-42} L(\rm H\alpha)/\lum$
     \citep{1998ARA&A..36..189K},
     the extended \ha\ emission requires merely a SFR of \rev{\mbox{0.01 \msun~yr$^{-1}$}}.

 %% Section 7
 \section{The scenario: tidal accretion and merger}

  \subsection{Creating the ring and extended \ha-emitting gas}

     %% According to the 2-dimensional decomposition for \thisobj, the color of the ring,
     %% $R - H$ = 3.34, is close to
     %% that of the bulge ($R - H$ = 3.55). This suggests the main content of the ring is
     %% the stellar component.
     %%%% the above is irrelevant . --- 2016Oct7 ----
     %%%%
     As mentioned in \S5,  the ring shows a disperse shape on its south side in the sky.
     We note that some parts of the inner region of the ring are
      filled by the galactic disk component
      %%% if you use "some parts of...", then you don't need to use `almost'.
     (see the \hst\ FW160 image in Figure~\ref{fig:2ddecompose}).
     %%In addition,
     Besides, the physical scale of the extended \ha\ emission is larger than the ring, and the bright spots in
     the extended \ha\ emission along the slit orientation
     are located interior to the ring.  %%within the inner region of the ring. -2016Nov21--%
     %%is just corresponding to the inner region of the ring.
     Plausibly, the extended \ha\ emission powered by {\it in situ} star formation,
     the faint galactic disk and the circumgalactic ring have a certain connection.

     %%According to modern scenarios,  -- 2016Oct7 ---
     The kind of galactic rings that are not associated with galactic bars,
     either coplanar or not with the disk of the host galaxies,
     are collectively called collisional rings or polar rings.
     They are generally believed to be
     formed by a violent interaction (collision or merger)
     and/or the tidal accretion during such an interaction between two galaxies
     \citep{1976ApJ...209..382L,1976ApJ...208..650T,1978IAUS...79..109T,2003A&A...401..817B,2012MNRAS.420.1158M}.
     Thus, the
     presence of the circumgalactic ring suggests that \thisobj, unbarred and dominated by the bulge,
     once experienced a recent galactic interaction.
     %%% -- 2016Oct7 ---
     %%%

     %% to start a new paragraph : 2016Nov21 : --- %%
     Here we propose a scenarios to explain the origin of the ring and the extended \ha-emitting gas,
     as well as the galactic disk perhaps,  as follows.
     %% --- now describe the scenario : ---- 2016Oct7 ---
     \thisobj\ has undergone a recent, violent interaction with an intruder galaxy.
     %% of a comparable mass and with rich gas.
     In the SDSS field of view around \thisobj, we do not find any a galaxy can be (the remnant of) the intruder;
     this is not like the usual situation of collisional ring events \citep[e.g.,][]{2016ApJ...822...64L}.
     Thus the intruder much have been instead merged with the host (see, e.g., \citealt{2003A&A...401..817B}),
     becoming part of the bulge as in common major mergers.
     During the interaction/collision stage,
     as the numerical simulations of \citet{2003A&A...401..817B} suggested,
     the gas of the intruder is tidally accreted by the host galaxy and forms the ring.
     According to their simulations,
     the intruder should be gas-rich and of a comparable mass to \thisobj.
     Then, in the merging stage, part of the stars of the intruder (or part of the stars of the host as well)
     is spread, forming the faint disk.
     Likewise, part of the gas of the intruder (or, secondarily, part of the gas of the host as well) is spread,
     which is the origin of the large-scale \ha-emitting gas.

     As for the rotation-like pattern of the velocity offsets of the extended \ha\ emission, there are two possible explanations.
     One possibility is that
      the ionized gas traces  the outflowing gas in galactic winds.
     The aforementioned violent galactic interaction triggers the star formation activities,
     which can produce
     galactic winds, as detected in various kinds of galaxies from ULIRGs to dwarf starburst galaxies
     (e.g., NGC1569; \citealt{2002ApJ...574..663M}).
     The outflowing gas might move outwards in close to perpendicular angles with respect to our line of sight,
     and thus the observed  large-scale \ha\
     emission would have only small blueshifts and redshifts.
     Yet galactic winds are
      seldom reported in galaxies with such red colors as \thisobj.
     %%% 2016Oct7 ---
     %% However, we do not find the remnant
     %% bullet galaxy nearby \thisobj\ in SDSS and \hst\ images.
      The other possibility (cf. \citealt{2003A&A...401..817B}), as we prefer, is that the extended \ha\ emission traces
     the large-scale gas that is formed by the tidal accretion/disruption from the intruder galaxy
     and that has an angular momentum as the rotation-like pattern suggests.
     The velocity profile of the extended \ha\ emission in \thisobj\ is similar to the large-scale rotation gaseous disks
      reported in the literature \citep[e.g.,][]{2011ApJ...740...83K}.
      Thus this explanation fits in well the whole scenario we propose above.

%% \S7.2  -- ? 2017Jan24
     \subsection{Triggering the AGN?}

     It is interesting to think about the following question: How is the IMBH AGN triggered in such a red, old and (presumably)
     gas-poor galaxy?
     First of all, we should agree that, as current observations indicate, the majority---likely more than about two third---of
     low-$z$ IMBH AGNs (and almost all the AGNs in nearby galaxies, \citealt{2008ARA&A..46..475H})
     are triggered by internal secular processes caused mainly by various instabilities of galactic disks
     and by nuclear star formation activities (\citealt{2008ApJ...688..159G,2011ApJ...742...68J};
     see \citealt{2013ARA&A..51..511K} for a detailed review).
     That is, major merger is {\em not the only way} to break the `angular momentum barrier' in the AGN fueling passage
     and trigger the AGN
     %in the final 1~pc of the  AGN fueling passage
     \citep{2006LNP...693..143J}.
     This is definitely the case for AGNs in local massive galaxies, because for galaxies of
     such masses ($\gtrsim L_{MW}$) in the local Universe
     the frequency of ongoing/recent major majors is indeed small \citep{2009ARA&A..47..159B},
     consistent with both the expectation of the hierarchical clustering theory and
     the observed ``cosmic downsizing'' phenomenon \citep{1996AJ....112..839C}
     that most star formation and most AGN activity \citep{2014ARA&A..52..589H}
     now happen in relatively low-mass galaxies.

     But, for low-mass galaxies in the local Universe, the major-merger frequency is much higher than the massive cousins.
     This is mainly because of the large number of low-mass galaxies, and secondarily because
     they are usually dragged by massive DM halos and thus close to each other, forming `groups' of low-mass galaxies.
     Both factors are just as the hierarchical clustering theory expects actually
     (see \S1; e.g., \citealt{2014ApJ...794..115D}).
   %%
   %%% 2016Sep23 evening ------
   % Numerical studies show that the statistical probability of a merger of galaxies decreases toward the low-mass regime
   % \citep[e.g.,][]{2006MNRAS.366..499D,2010MNRAS.401.2245F}.
   % Having a shallow potential well, low-mass galaxies are thought to have an evolution that is more
   % driven by so called ``nature'' and ``nurture'', which indicate the internal effects (e.g., total mass)
   % and the large-scale environment respectively, rather than
   % major merger events \citep[e.g.,][]{2012MNRAS.420.1714S}.
   %
   %As to the low-mass galaxies with AGN activities,  \citet{2008ApJ...688..159G,2011ApJ...742...68J} show that
   % almost all low-mass BHs live in low-mass, Sbc--Sd spirals that contain pseudobulges,
   % and 1/3 of the host galaxies of low-mass BHs contain bars.
   % Based on these, \citet{2013ARA&A..51..511K} concludes that mergers are not important
   % for these low-mass active galaxies,
   %and the these BHs grow mainly by secular processes.
   %%% -- 2016Sep23 evening ------
   %
   %Recent numerical studies pay more attention to the role of dwarf-dwarf major mergers/interactions in low-mass galaxy evolution.
   %\citet{2014ApJ...794..115D} investigate the frequency of major mergers between dwarf galaxies in the Local Group. The find that
   %$\sim 10$\% of satellite dwarf galaxies with $M_{\rm star} > 10^{6}$\msun\ that are within the host virial radius experienced a
   %major merger of stellar mass ratio closer than 0.1 since $z = 0.1$.
   %% --- the above is already in the Introduction ----. %%
   %%
   Recently, reported observational cases of dwarf-dwarf major mergers also have increased
   \citep[e.g.,][]{2005AJ....129.2617G,2013ApJ...779L..15N,2014ApJ...795L..35C,2014Natur.507..335A,2015AJ....149..114P}.
   Based on a sample of 104 dwarf galaxy pairs, \citet{2015ApJ...805....2S} show that the star formation enhancement observed
   in massive galaxy pairs extends to the dwarf mass range also.
   Distorted \hi\ morphologies and the presence of gaseous tails around star-bursting dwarf galaxies have also been attributed to
   wet major mergers \citep{2013ApJ...779L..15N,2016MNRAS.459.1827P}.
   Again, all these observations are not inconsistent with the observed ``cosmic downsizing'',
   but raise a long neglected phenomenon---wet major mergers/violent interactions between dwarf galaxies and their same effect onto
   star formation as their massive cousins did in their active epochs (at higher redshifts).

    %% a new paragraph: dwarf-dwarf major mergers to trigger AGN ------
   We can suppose that, like the aforementioned role in triggering star formation,
    wet major mergers/violent interactions of local gas-rich small galaxies play almost the same role in triggering AGN activity as
   their massive cousins plausibly did at high redshifts (e.g., $z\approx2-3$).
   The high-$z$ role
   is commonly believed in the co-evolution paradigm of supermassive BHs and their massive host galaxies
   in their quasar epoch \citep{2013ARA&A..51..511K}.
   Then our proposed scenario for the ring and extended \ha-emitting gas can naturally explain
   the triggering of the AGN too.
   %%% added on 2017Jan24 below: ----
\rev{%%2017Jan:
   Certainly, we should bear in mind a second possibility: the co-occurrence of the recent major merger event
   and the AGN activity in \thisobj\ is just accidental, because neither the major-merger rate nor the AGN fraction
   is small in local low-mass galaxies.
   The conclusion cannot be drawn until well-defined comparisons are done in major-merger rate
   between galaxies with AGN activities and the same kind of galaxies without.
}

   What we would like to stress again is that
   %% the role  -- 2016Sep7: not 'the role', but the incidence -- %%
   the incidence of major mergers and other violent interactions in local IMBH AGNs
   is not insignificant at all.
   According to the studies mentioned in \S1 (\citealt{2011ApJ...742...68J}, W.-J.~Liu \etal\ 2017, in preparation),
   the fraction of such violent galaxy interactions in IMBH host galaxies is about 9--16\%,
   which is much higher than in local massive galaxies.
   In fact, the above values are
   similar to the fraction of the radio-loud population in AGNs and to the fraction of broad-absorption-line (BAL) quasars;
   both radio-loud AGNs and BAL quasars have not been neglected for their not being the majority.
   We think that violent galaxy interactions in IMBH AGNs are also worth further investigations.

     \section{Summary and future work}

     %% its peculiarity --
     \thisobj\ first caught our attention in the \citet{2012ApJ...755..167D} sample of 309 IMBH AGNs,
     as it is located at an extreme portion of the parameter space with certain peculiarities:
     host galaxy being rather red ($u-r = 2.41$) for a total luminosity \mbox{$M_V = -17.80$ mag},
     and AGN belonging to the ten LINER 1s in the sample.
     %%
     %% From its SDSS spectrum, the continuum of \thisobj\ is dominated by host-galaxy starlight.
     %% Optical emission lines reveal
     %% the nucleus to be a broad-line LINER.
     %%% 2016Nov22 --- .
     In the present work, we carried out a detailed study with its \hst\ images and
     MMT longslit spectroscopic data as well as the SDSS data.
     The analysis of the \hst\ images
     indicates that \thisobj\ is an early-type galaxy:
     the disk is featureless, faint, and of low surface brightness (\mbox{$\mu_{0,V} = 20.39$\,mag\,arcsec$^{-2}$})
     whereas the bulge is prominent, red ($V- I =2.03$), relatively compact ($R_{\rm e} = 0.28$\,kpc)
     and accounts for $\approx81$\% of the total light in the $I$ band.
     Interestingly,  the \hst\ images reveal a circumgalactic ring with a 16~kpc diameter.
     The ring is clear-cut and narrow on its north side but manifests a dispersed shape on its south side.
     More surprisingly, the MMT 2-dimensional longslit spectrum reveals
     large-scale \ha\ emission with a projected physical size of 20~kpc.
     The velocity profile of the extended \ha\ shows a rotation-like pattern, with a mean blueshifted velocity of \mbox{162\,\kms}
     and mean redshifted velocity of \mbox{86\,\kms}.
     %% 2017Feb3:
     %% Our photoionization calculation suggests that the large-scale \ha-emitting gas is unlikely to be powered by
     %% direct photoionization of the central nucleus.
Our photoionization calculations suggest that 
the AGN continuum powers the broad \ha\ emission line, 
the post-AGB stars power (part of) 
the normal-aperture narrow emission lines as indicated by the LINER-like narrow-line spectrum in terms of the BPT diagram,
but neither is sufficient to further power the large-scale \ha\ emission. 
%%%2017Feb3. ---     
     This is vindicated by the BPT diagnostic of the extended emission,
     which also disfavors any a kind of shock heating but indicates a star formation origin.
     We propose that both the ring and extended \ha-emitting gas are created by the tidal accretion
     in a collision event (and then merger) with a gas-rich galaxy of a comparable mass.
     Provided that wet major mergers (or other violent interactions) of {\em present-day} small galaxies
     play almost the same role in triggering AGN activity as
     their massive cousins supposedly did at high redshifts (namely the co-evolution of high-$z$ quasars and massive galaxies),
     then our proposed scenario can naturally explain
     the triggering of the IMBH AGN too.
     In summary, \thisobj\ is a good target to study the influence of violent galactic interactions on IMBH AGNs (plausibly) and
     on their (dwarf) host galaxies.

    %%% a new paragraph : the future work --- 2016NOv22: --- %%%
     We are proposing deep and high-spatial-resolution observations in multiple bands to further investigate
     the concrete astrophysical processes at play in  \thisobj.
     IFU observations in the optical and near-infrared, particularly with AO-assisted,
     will obtain the spatially resolved kinematics of
     the host-galaxy stars, and the spatial distribution and kinematics of both the star formation and the large-scale warm gas.
     Radio aperture-synthesis observations of high spatial resolution (e.g., by JVLA, GMRT, and ALMA in particular)
     will reveals the spatial distribution and kinematics of the cold gas.
     Those observations will enable us to probe the spatial scales from the circumgalactic down to
     tens of parsecs from the nucleus, with a panoramic view.
     Meanwhile, as a straight step forward, it is exciting to find more systems like \thisobj\ and
     observationally census the frequency of violent galactic interactions in the dwarf-galaxy world generally,
     and in the dwarf-AGN world particularly.

%\\ acknowledgement
\vspace{6pt}
%% ---  to add the thanks later .... Oct7 ---
   We thank the anonymous referee for the thoughtful suggestions and comments with great expertise 
   in spectral and imaging handlings that improved this paper significantly,
   and thank Chien Y. Peng, Hongyan Zhou, Weimin~Gu, and Shaohua Zhang for their helpful discussions.
   This work is supported by
%%% #1:
   National Basic Research Program of China (973 program No. 2015CB857101), %%L.Qian--xao-973
%%% #2:
   the International Partnership Program of Chinese Academy of Sciences (Grant No. 114A11KYSB20160008), %%Di Li
%%% #3:
   Natural Science Foundation of China grants (NSFC 11473062, %%dxb-mianshang
   11603036, %%L.Qian-qingnian-2016
   and 11603021), %%JN-2016
%%%
   and the fundings from the
   Key Laboratory for the Structure and Evolution of Celestial Objects of Chinese Academy of Sciences
   (Grant No. OP201408) and the CAS Interdisciplinary Innovation Team.
%%%
   QL is supported by the FAST Fellowship. The FAST Fellowship is supported by Special Funding for Advanced Users,
   budgeted and administrated by Center for Astronomical Mega-Science, Chinese Academy of Sciences (CAMS).
   This work has made use of the data products of the SDSS and \hst, and data taken by the MMT telescope
   through the Telescope Access Program (TAP).
   TAP is funded by the Strategic Priority Research Program ``The Emergence of Cosmological Structures''
   (Grant No. XDB09000000), National Astronomical Observatories, Chinese Academy of Sciences, and the Special
   Fund for Astronomy from the Ministry of Finance.
   The MMT Observatory is a joint facility of the Smithsonian Institution and the University of Arizona.

\clearpage

 %%%% Figure 1
\begin{figure}[htbp]
       \center
       \includegraphics[width=6.0in]{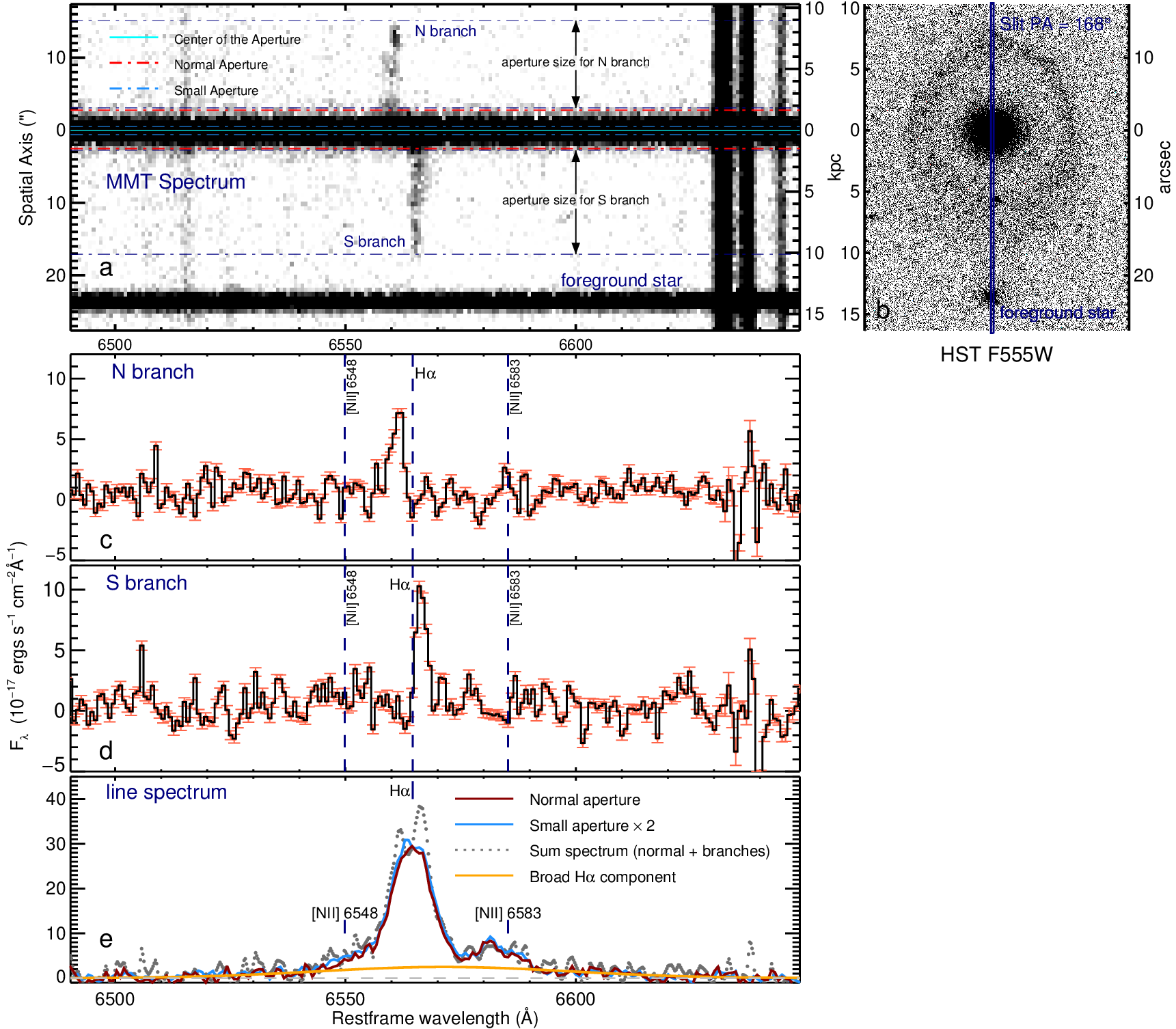}
       \caption{{\it Panel a}: Two-dimensional longslit spectrum observed with MMT Red Channel.
         The x-axis is the restframe vacuum wavelength along the dispersion direction.
         The y-axis is the spatial axis along the slit;
         on the left  angular scale is  labeled relative to
         the spatial aperture center (cyan solid line),
         and on the right, the projected physical scale.
         Dash-dotted lines of different colors denote the different spatial apertures used to extract the spectrum;
     see the text for the detail.
     %The red lines are
     %for normal aperture, the blue lines are for the small aperture and the navy-blue lines show
     % the ranges of aperture size to extract of spectra of branch N. and branch S.
         {\it Panel b}: \hst\ F555W image, with the same spatial scale as the 2-dimensional spectrum in the $y-$axis.
         A foreground star to the south is denoted. The MMT slit is oriented along the $y-$axis, through the centers of \thisobj\
         and the denoted foreground star.
         {\it Panel c}: One-dimensional spectrum of the N branch.
         {\it Panel d}: One-dimensional spectrum of the S branch.
         {\it Panel e}: Starlight-subtracted spectra of different extracted apertures.
     The small-aperture spectrum is scaled to the flux level of the normal-aperture one.
     The sum spectrum
     (normal-aperture $+$ N~branch $+$ S~branch; the gray dotted line)
     is plotted for comparison.
     Also plotted is the best-fit broad~\ha\ component
     (the orange solid line; see Figure~\ref{fig:decompose}).
     \label{fig:MMTimage}}
  \end{figure}

% %%%% Figure 2
\begin{figure}[htbp]
       \center
       \includegraphics[width=6.0in]{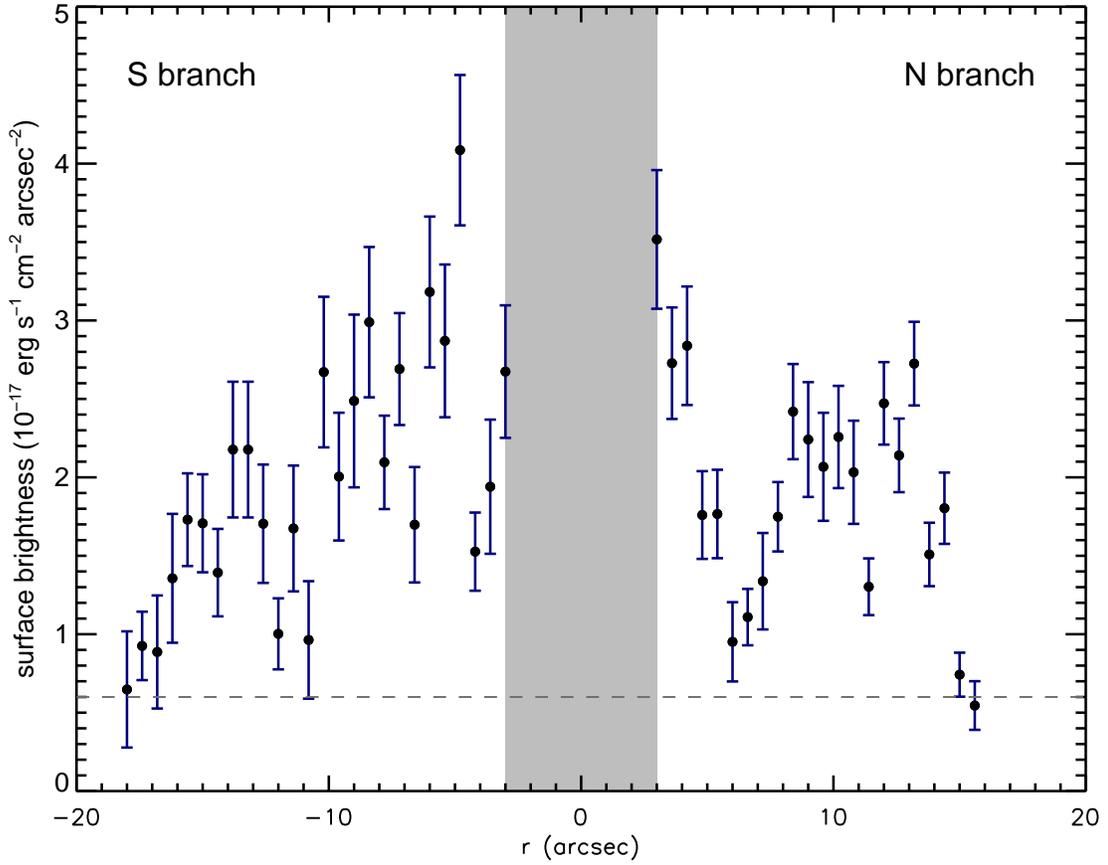}
       \caption{\label{fig:ha_radial}
The radial profile of the surface brightness of the extended \ha\ emission.
      North is toward positive values in the spatial axis $r$.
      The gray shaded area denotes the region affected by \ha\ emission from the AGN BLR and NLR.
The dashed line  denotes the lowest surface brightness of the extended \ha\ emission
we can probe, in terms of which
the outer bounds of the extraction apertures are set for the two branches.
       }
  \end{figure}

%%%%%% Figure 3

  \begin {figure}[htbp]
      \centering
      \includegraphics[width=6.2in]{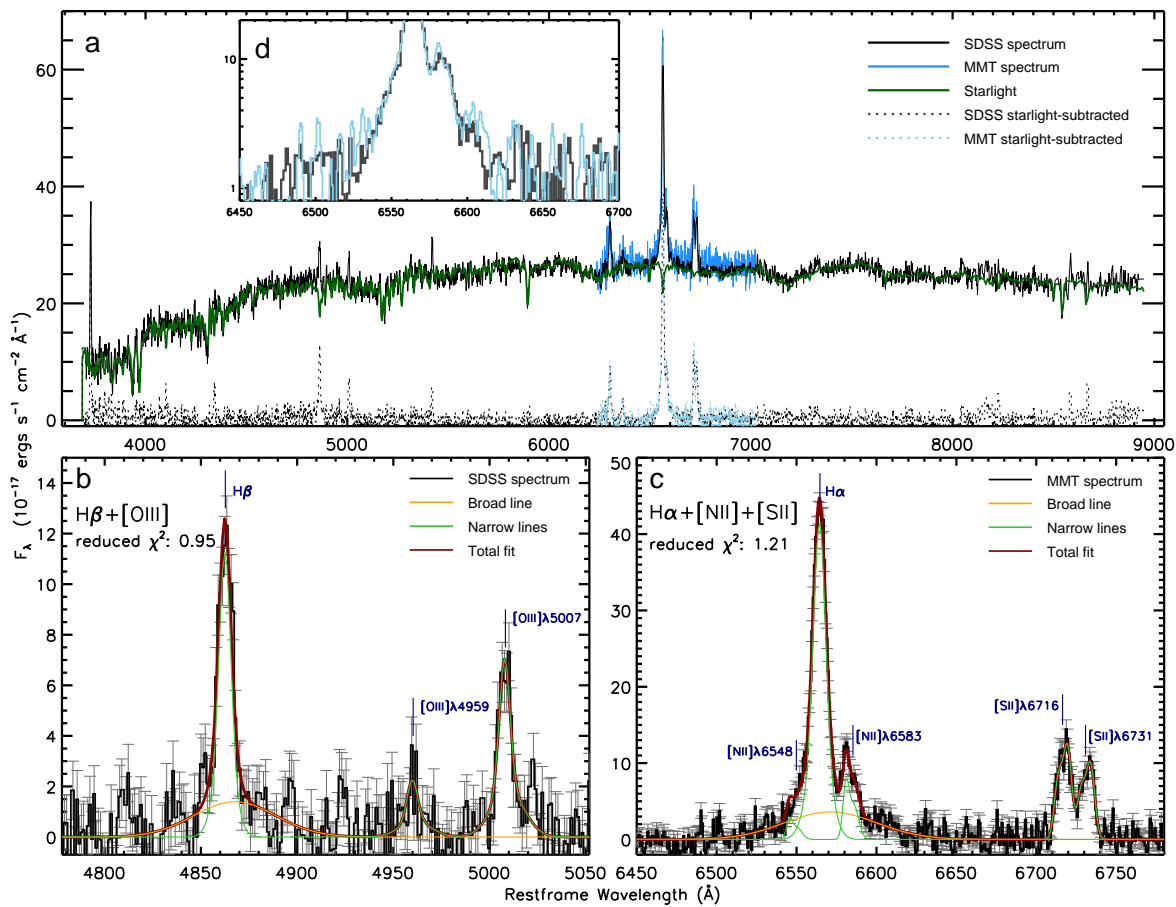}
      \caption{\footnotesize SDSS and MMT spectra, together with the continuum decomposition and line fitting.
    {\it Panel a}: The SDSS (black) and MMT (blue) spectra,
    the best-fit starlight continuum (green), and the respective starlight-subtracted spectra.
    The MMT spectrum
    is extracted with the ``normal aperture'' showed in Figure~\ref{fig:MMTimage},
    and scaled to the SDSS one (see \S2.2).
    {\it Panel b}: The best-fit model for the emission lines in the \hb$+$\oiii\ region.
    {\it Panel c}: The best-fit model for the emission lines in the \ha$+$\nii$+$\sii\ region.
    {\it Panel d}: Comparing the starlight-subtracted spectra: the SDSS (black) versus the MMT (blue).
    \label{fig:decompose}}
  \end{figure}

%
%
%
%  %---------------------------------
%
%% Table 1
 \begin{deluxetable}{llcc}
 \tablecolumns{4}
 \tablewidth{0pc}
 \tabletypesize{\footnotesize}
 \tablecaption{Emission-Line Parameters  \label{tab:linepar}}

 \tablehead{
   \colhead{Emission Line} & \colhead{Centroid\tablenotemark{a}} & \colhead{FWHM\tablenotemark{b}} & \colhead{Flux} \\
   \colhead{}              & \colhead{\AA}      & \colhead{\kms} & \colhead{10$^{-17}$ erg s$^{-1}$ cm$^{-2}$}}
 \startdata
 \ [O\,II]$\lambda3727$                               & \rev{3728.15$\pm$0.19} & \rev{528$\pm$46  } & \rev{ 217$\pm$14} \\
  \ H$\beta\lambda4861$(broad)\tablenotemark{c}       & \rev{4867.07         } & \rev{2814        } & \rev{  81$\pm$13} \\
  \ H$\beta\lambda4861$(narrow)\tablenotemark{d}      & \rev{4862.42         } & \rev{475         } & \rev{  93$\pm$5 } \\
  \ [O\,III]$\lambda5007$                             & \rev{5008.64$\pm$0.42} & \rev{474$\pm$73  } & \rev{  75$\pm$8 } \\
  \ [O\,I]$\lambda6300$                               & \rev{6300.95$\pm$0.37} & \rev{548$\pm$35  } & \rev{ 111$\pm$5 } \\
  \ [O\,I]$\lambda6364$\tablenotemark{e}              & \rev{6364.44         } & \rev{548         } & \rev{  43$\pm$4 } \\
  \ H$\alpha\lambda6563$(broad)                       & \rev{6570.53$\pm$1.22} & \rev{2814$\pm$210} & \rev{ 286$\pm$22} \\
  \ H$\alpha\lambda6563$(narrow)                      & \rev{6564.21$\pm$0.07} & \rev{ 476$\pm$10 } & \rev{ 438$\pm$13} \\
  \ H$\alpha\lambda6563$(N branch)                    & \rev{6561.07$\pm$1.00} & \rev{ 162$\pm$13 } & \rev{ 28$\pm$0.8} \\
  \ H$\alpha\lambda6563$(S branch)                    & \rev{6566.49$\pm$1.00} & \rev{ 128$\pm$9  } & \rev{ 33$\pm$0.8} \\
  \ [N\,II]$\lambda6583$                              & \rev{6583.04$\pm$0.52} & \rev{469$\pm$48  } & \rev{  68$\pm$8 } \\
  \ [S\,II]$\lambda6716$                              & \rev{6717.36$\pm$0.54} & \rev{465$\pm$37  } & \rev{ 127$\pm$4 } \\
  \ [S\,II]$\lambda6731$\tablenotemark{f}             & \rev{6731.74         } & \rev{465         } & \rev{ 103$\pm$4 }
 \enddata
 \tablenotetext{a}{\footnotesize Vacuum restframe wavelength.}
 \tablenotetext{b}{\footnotesize Corrected for instrumental broadening.}
 \tablenotetext{c}{\footnotesize Adopting the profile of the broad component of H$\alpha$.}
 \tablenotetext{d}{\footnotesize Adopting the profile of the narrow component of H$\alpha$.}
 \tablenotetext{e}{\footnotesize Adopting the profile of [O\,I]$\lambda$6300.}
 \tablenotetext{f}{\footnotesize Adopting the profile of [S\,II]$\lambda$6716.}
 \end{deluxetable}

%%%%%% Table 2
 \begin{table}[htbp]
% \begin{sidewaystable}
   \begin{threeparttable}
   \centering
   \scriptsize
     \caption{Results of GALFIT decomposition}
     \label{tab:galfit}

     \begin{tabular}{ c  c  c c c  c  c  c  c c c c c c c}
       %\hline \hline
       \toprule \toprule
       %& \multicolumn{2}{c}{Nonparametric} &  & \multicolumn{10}{c}{Parametric}  \\
       %\cline{2-3}\cline{5-14}\\
       & \multicolumn{2}{c}{PSF}  &   & \multicolumn{4}{c}{Bulge} &  &  \multicolumn{3}{c}{Disk}&   &  \multicolumn{2}{c}{Ring}    \\
       \cline{2-3}\cline{5-8}\cline{10-12}\cline{14-15}\\
       Band & $m$  & $M$ & &  $n$  & $m$  & $M$  & $r_e$          & &$m$ & $M$ & $r_s$          & & $m$  & $M$    \\
            &      &     & &       &      &      & (\arcsec/kpc)& &    &     & (\arcsec/kpc)& &      &        \\
       (1)  & (2)  & (3) & &  (4)  & (5)  & (6)  &  (7)           & &(8) & (9) & (10)           & & (11) & (12)   \\
       %\hline
       \midrule
       \hst\ F555W                   & 21.45 & $-$14.11 &  & 1.83 & 18.61  & $-$16.95 &0.48/0.28 &  & 18.52 & $-$17.04 & 0.92/0.54 & & 21.21     & $-$14.35 \\
       \hst\ F814W$^{\,\rm a}$ & 21.00 & $-$14.56 &  & 4.00 & 16.58  & $-$18.98 &0.93/0.55 &  & 18.26 & $-$17.30 & 1.03/0.61 & & \nodata  &  \nodata      \\
       \hst\ F160W                   & 18.83 &  $-$16.73 &  & 3.29 & 15.07 & $-$20.49 &0.63/0.37 &  & 16.97 & $-$18.59 & 1.13/0.67 & & 17.91     & $-$17.65  \\
       %\hline
       \bottomrule
      \end{tabular}
  \begin{tablenotes}
  \item Col (1):    HST filter.
  \item Col (2):    the integrated magnitudes on the Vega system for PSF component, not corrected for Galactic extinction.
  \item Col (3):    the absolute magnitude for PSF component after Galactic extinction correction.
  \item Col (4):    the S\'{e}rsic index.
  \item Col (5):    the integrated magnitudes on the Vega system for bulge component, not corrected for Galactic extinction.
  \item Col (6):    the absolute magnitude for S\'{e}rsic component after Galactic extinction correction.
  \item Col (7):    the effective radius of the S\'{e}rsic component in unit of arcseconds and the corresponding scale length
                    in unit of kpc.
  \item Col (8):    the integrated magnitudes on the Vega system for disk component, not corrected for Galactic extinction.
  \item Col (9):    the absolute magnitude for disk component after Galactic extinction correction.
  \item Col (10):   the effective radius of the disk component in unit of arcseconds and the corresponding scale length
                    in unit of kpc.
  \item Col (11):   the integrated magnitudes on the Vega system for the outer ring, not corrected for Galactic extinction. The
    ring magnitude is derived from the residual images.
  \item Col (12):   the absolute magnitude for the outer ring component after Galactic extinction correction.
  \item $^{\rm a}$   The decomposition result of \hst\ F814W is taken from \citet{2011ApJ...742...68J}.
  \end{tablenotes}
\end{threeparttable}
\end{table}
% \end{sidewaystable}

%% Figure 4

  \begin{figure}[htbp]
      \centering
      \includegraphics[width=5.0in]{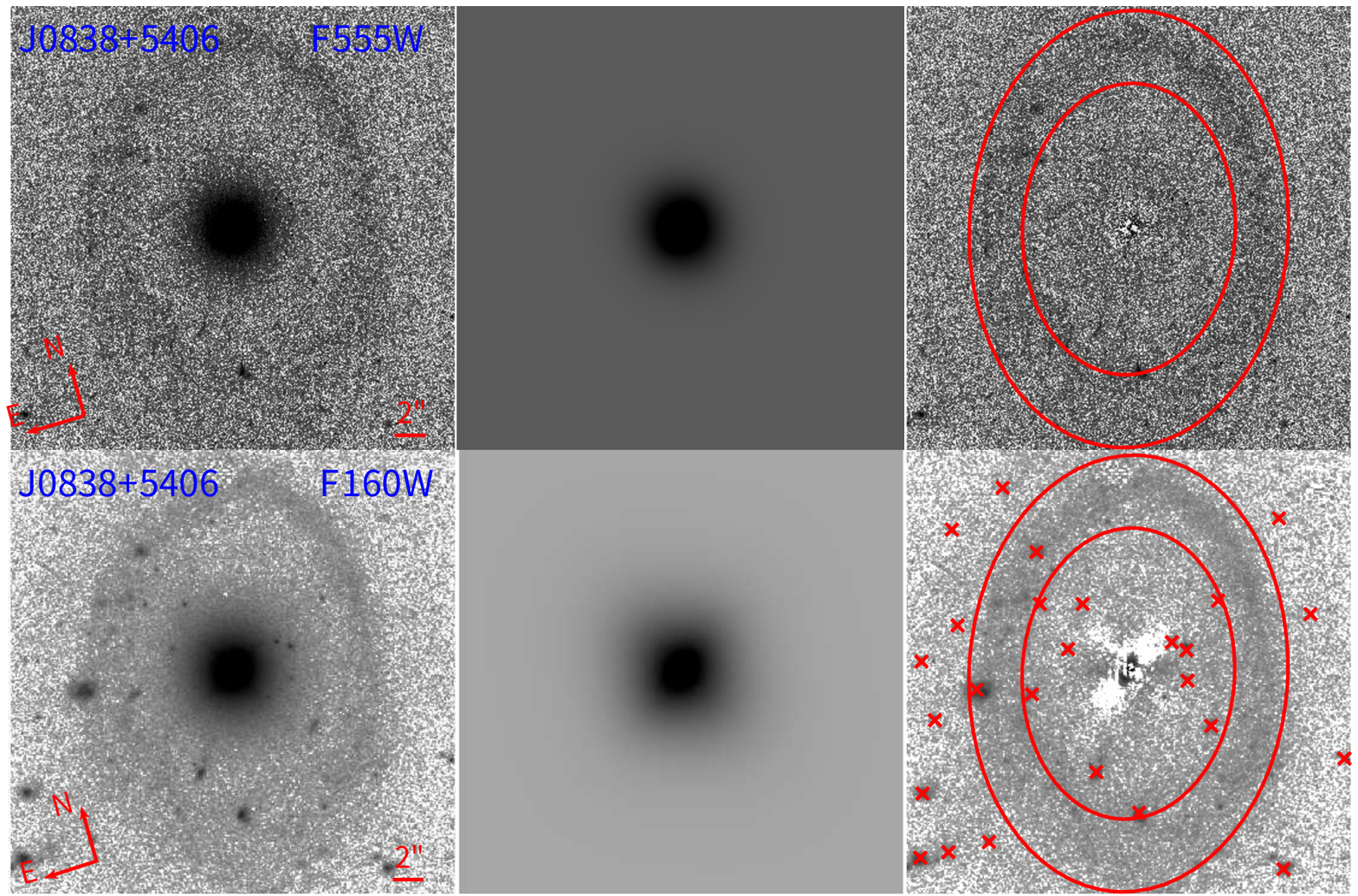}
      \includegraphics[width=5.0in]{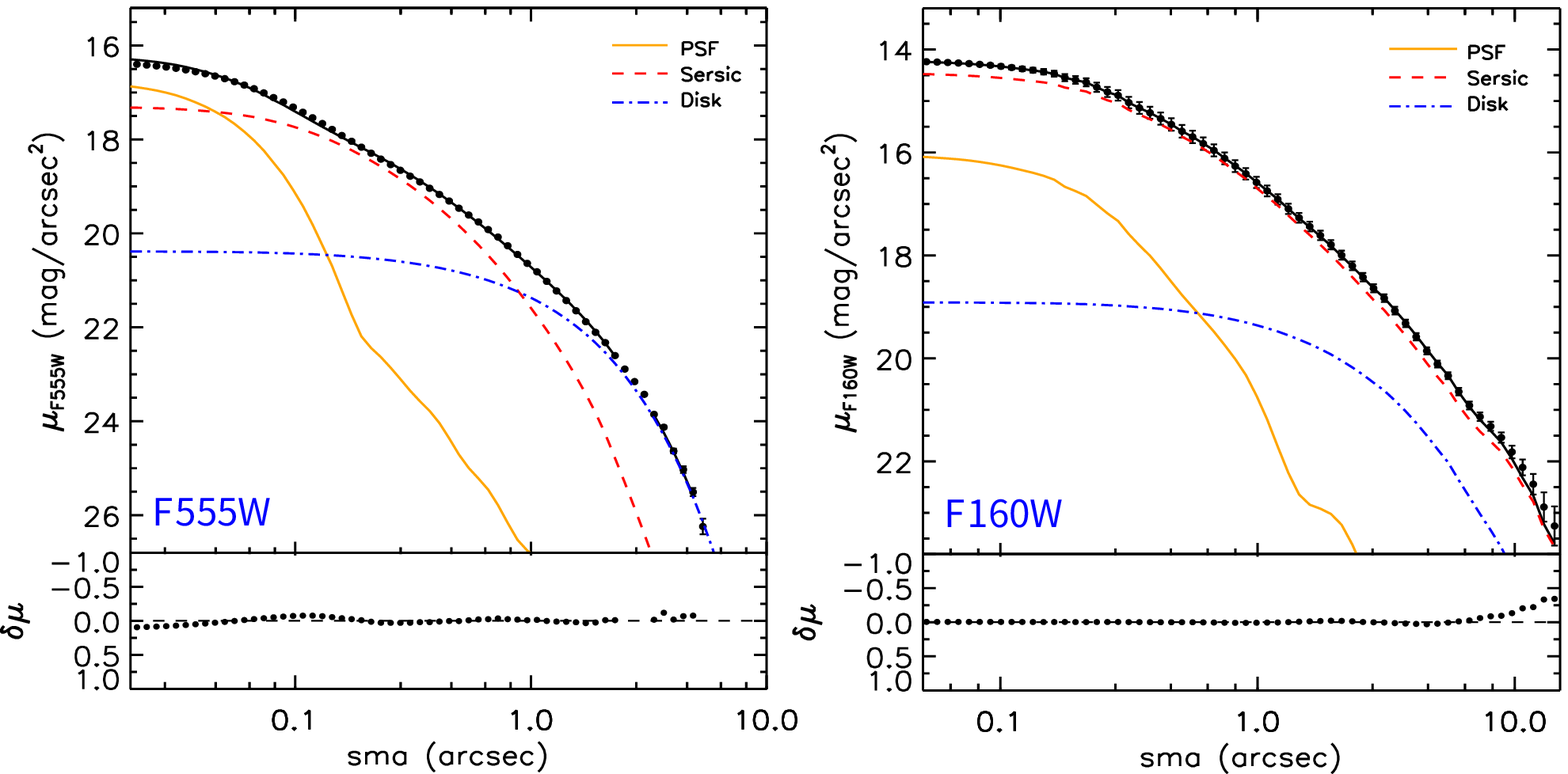}
      \caption{GALFIT decomposition of {\it HST}/WFC3 images of \thisobj.
      {\it Top panels}: The original, model and residual images from left to right,
       for the F555W band. The ring region is denoted by the annulus between the two red ellipses.
      During the fitting the ring region and other contaminations (marked as red crosses in the image blow)
      are masked out. %%2016Nov26%%
      {\it Middle panels}: The same but for the F160W band.
      {\it Bottom panels}:  One-dimensional representation of
      the three-component 2-D GALFIT model applied to the F555W (left) and F160W (right) images:
      PSF for the AGN (orange line), S\'{e}rsic function for the bulge
      (red dashed line), and an exponential function for the galactic disk (blue dot-dashed line).
      In every panel, the sum of the three components is shown as the black solid line;
      the observed data are plotted as black symbols with $\pm1\sigma$ error bars;
      shown in the bottom are
      the residuals between the data and the best-fit model.
      \label{fig:2ddecompose}}
  \end{figure}

%%% Figure 5:
  \begin{figure}[htbp]
      \centering
      \includegraphics[width=4.8in]{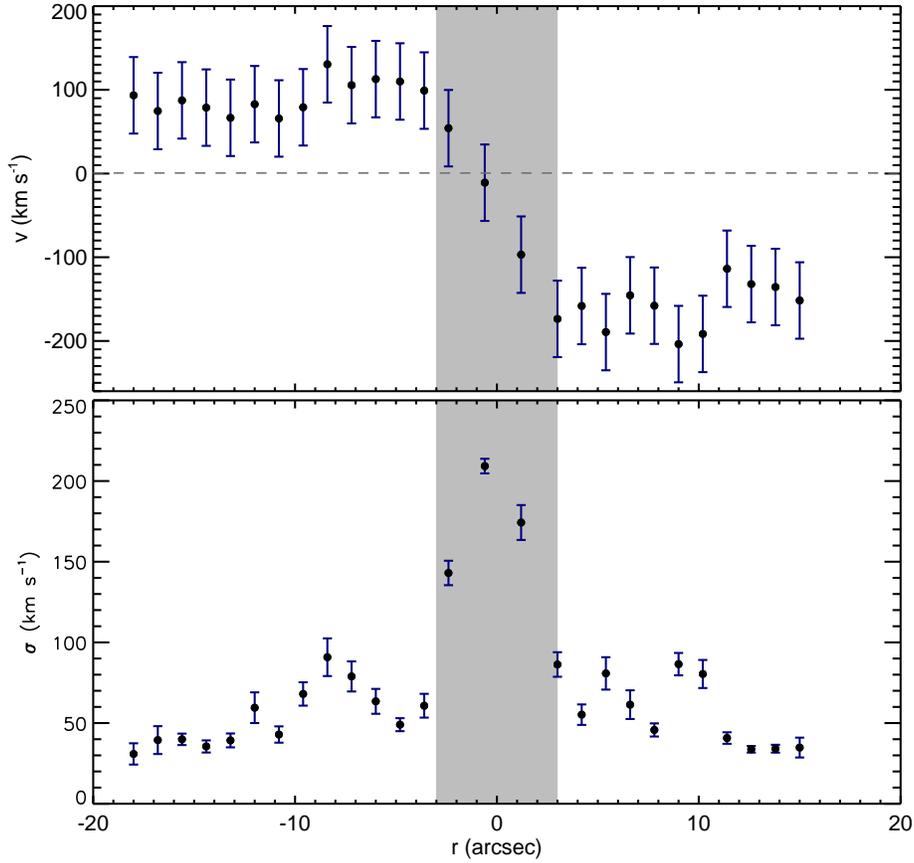}
      \caption{The mean ($v$, with respect to the systematic redshift) and
       dispersion ($\sigma$) of the line-of-sight velocity distribution of the \ha\ emission
       at different spatial positions along the slit,
       measured from the 2-D MMT spectroscopic data.
      The positive $v$ values denote red shifts, and the negative, blue shifts;
      the $\sigma$ values are not corrected for instrumental broadening ($\sigma_{\rm ins} \approx 36.5$\,\kms). 
      The spatial axis $r$
      and the gray shaded area
      are the same as in Figure~\ref{fig:ha_radial},
      but every 2 or 3 spatial pixels are combined to increase the S/N of the data points (see the text).
\label{fig:rotationcurve}}
  \end{figure}

\end{document}